\documentclass[10pt,journal,compsoc]{IEEEtran}
\ifCLASSOPTIONcompsoc
  \usepackage[nocompress]{cite}
\else
  \usepackage{cite}
\fi
\ifCLASSINFOpdf
\else
\fi
\usepackage{blindtext}
\usepackage{amsmath,amssymb,amsfonts}
\usepackage{graphicx}
\usepackage{xcolor}
\usepackage{makecell}
\usepackage{subfigure}
\usepackage{xspace}
\usepackage{amsthm}
\usepackage[ruled,linesnumbered]{algorithm2e}
\usepackage{algpseudocode}
\usepackage{comment}
\usepackage{caption}

\newcommand*{\QEDA}{\null\nobreak\hfill\ensuremath{\blacksquare}}%
\newcommand*{\defeq}{\stackrel{\text{def}}{=}}
\newcommand{\BfPara}[1]{\vspace{1mm}{\noindent\textbf{#1:}}\xspace}
\newcommand{\etal}{{\em et al.}\xspace}
\newcommand{\ours}{Count-Less}
\newcommand{\cmin}{Count-Min}
\newcommand{\cmcu}{CM-CU}
\newcommand{\clcm}{CL-CM}
\newcommand{\clcu}{CL-CU}
\newcommand{\clmu}{CL-MU}
\newcommand\eg{e.g.,\xspace}
\newcommand\ie{i.e.,\xspace}

\newcommand{\sy}[1]{{\textcolor{teal}{#1}}}

\definecolor{LightGray}{gray}{0.98}
\theoremstyle{plain}
\newtheorem{theorem}{Theorem}

\begin{document}
%
\title{\ours{}: A Counting Sketch for the Data Plane of High Speed Switches}
%
%
%
%


\author{SunYoung~Kim,
        Changhun~Jung,
        RhongHo~Jang,
        David~Mohaisen,
        and~DaeHun~Nyang
\IEEEcompsocitemizethanks{\IEEEcompsocthanksitem S. Kim is with the Department of Cyber Security, Ewha Womans University, Seoul, Korea.
C. Jung is with the Department of Electrical and Computer Engineering at Inha University, Incheon, Korea.
R. Jang is with the Department of Computer Science, Wayne State University, Michigan, USA.
D. Mohaisen is with the Department of Computer Science, University of Central Florida, Florida, USA.
D. Nyang is with the Department of Cyber Security, Ewha Womans University, Seoul, Korea.
D. Nyang (nyang@ewha.ac.kr) is the corresponding author.

}
}

\IEEEtitleabstractindextext{%

\begin{abstract}
Demands are increasing to measure per-flow statistics in the data plane of high-speed switches. Measuring flows with exact counting is infeasible due to processing and memory constraints, but a sketch is a promising candidate for collecting approximately per-flow statistics in data plane in real-time. Among them, 
Count-Min sketch is a versatile tool to measure spectral density of high volume data using a small amount of memory and low processing overhead. Due to its simplicity and versatility, Count-Min sketch and its variants have been adopted in many works as a stand alone or even as a supporting measurement tool. However, Count-Min's estimation accuracy is limited owing to its data structure not fully accommodating Zipfian distribution and the indiscriminate update algorithm without considering a counter value. 
This in turn degrades the accuracy of heavy hitter, heavy changer, cardinality, and entropy. To enhance measurement accuracy of Count-Min, there have been many and various attempts. One of the most notable approaches is to cascade multiple sketches in a sequential manner so that either mouse or elephant flows should be filtered to separate elephants from mouse flows such as Elastic sketch (an elephant filter leveraging TCAM + Count-Min) and FCM sketch (Count-Min-based layered mouse filters). 
In this paper, we first show that these cascaded filtering approaches adopting a Pyramid-shaped data structure (allocating more counters for mouse flows) still suffer from under-utilization of memory, which gives us a room for better estimation. To this end, we are facing two challenges: one is (a) how to make Count-Min's data structure accommodate more effectively Zipfian distribution, and the other is (b) how to make update and query work without delaying packet processing in the switch's data plane. Count-Less adopts a different combination of data structure and update algorithm, namely {\em Split Counter} and {\em Minimum Update} to resolve (a) and (b), respectively. Split Counter strategy is for Zipfian distribution, and Minimum Update is for cross-layer update like Count-Min, while minimizing the number of updated counters and enabling the update with one time access of registers in data plane. Unlike the multi-sketch cascaded counters such as FCM and Elastic sketch, Count-Less is a single sketch of Count-Min type.  
Count-Less is much more accurate in all measurement tasks than Count-Min and outperforms FCM sketch and Elastic sketch, state-of-the-art algorithms without the help of any special hardware like TCAM. Not only theoretical proof on Count-Less's estimation but also comprehensive experimental result are presented in terms of estimation accuracy and throughput of Count-Less, compared to Count-Min (baseline), Elastic sketch, and FCM sketch. To prove its feasibility in the data plane of a high-speed switch, Count-Less prototype on an ASIC-based programmable switch (Tofino) is implemented in P4 language and evaluated. In terms of data plane latency, Count-Less is 1.53x faster than FCM, while consuming much less resources such as hash bits, SRAM, and ALU of a programmable router.
\end{abstract}

}

\maketitle

\IEEEdisplaynontitleabstractindextext

%
\IEEEpeerreviewmaketitle

\IEEEraisesectionheading{\section{Introduction}\label{sec:intro}}

%
%
%
%
\IEEEPARstart{T}{o} perform network monitoring and attack detection functions, a fine-grain per-flow traffic measurement is essential. However, one of the most extraordinary challenges in this space has been designing accurate algorithms for per-flow traffic measurement in the data plane of high speed switches. To date, various approaches have been proposed to address this challenge, including hardware-based approaches, sampling-based approaches, and sketch-based approaches. Recently, the networking community has been moving towards ASIC-based programmable switches to reduce the control loop, although such approaches still face several challenges, mainly due to the constraints of hardware designs. Especially hardware-based approaches leveraging the parallel search capability of TCAMs have been proposed, but they suffered from scalability issues owing to the scarcity of TCAM~\cite{BGPstat}.

Driven by the ever-increasing traffic volumes, network devices' bandwidth is also increasing dramatically, requiring scalable measurement functions not to affect the latency of the packet processing pipeline. This in turn made the sampling-based approaches a viable option due to their simplicity and scalability. However, it is often criticized for the poor trade-off between accuracy and overhead according to the sample rate. Demands of better sampling strategy for better accuracy with lower overhead motivated several recent works \cite{YHS21self, jang2020sketchflow,kumar2006sketch, liu2019nitrosketch}.

Sketch-based approaches are yet another promising option for per-flow measurement. Since Morris introduced the sketch concept in 1978~\cite{Morris78a}, and Flajolet \etal presented the FM-sketch in 1985~\cite{flajolet1985probabilistic}, a large body of works have been proposed in this space, including~\cite{cormode2005improved,kumar2004data, estan2002new,Whang:1990, nyang16,jang2020sketchflow, kumar2006sketch, liu2019nitrosketch, zhou2018cold, YangZJCL17Pyramid,LuMPDK08CounterBraids, song2020FCM}. 
In general, a sketch is evaluated by its estimation accuracy under processing and memory constraints. The lighter encoding/decoding algorithms are, the higher the bandwidth they can support. 

Among many interesting sketches, \cmin{} sketch \cite{cormode2005improved} is a simple yet powerful sketch that approximately counts a large volume of data stream using a small amount of memory. The simplicity of its operations (\ie encoding and decoding) and data structure have led to its use in multiple application domains, such as network measurement~\cite{huang2017sketchvisor, yang2018elastic}, database~\cite{tirmazi2020cheetah}, security~\cite{zhang2020poseidon}, natural language processing applications~\cite{goyal2011approximate}, and storage~\cite{goswami2018buffered}, among many others. Also, its estimation accuracy is good even with its over-estimation, which is controllable with proper parameter choice. Range, heavy hitter and quantile queries on the sketch are also supported, which makes \cmin{} attractive in many applications and network traffic measurement settings. Several extended versions of Count-Min sketch have been proposed to reduce errors and bias, such as Count-Mean-Min (CMM) ~\cite{deng2007new}, Count-Min sketch with Conservative Update (\cmcu{}) ~\cite{goyal2012sketch}, etc.
Due to its simplicity and versatility, \cmin{} sketch and its variants have been adopted in many works as a standalone or even as a supporting measurement tool~\cite{yu2013software, yang2018elastic}. \cmin{}'s measurement accuracy and speed determine the measurement system's performance, and thus the sketch's performance is paramount.

Despite its merits, \cmin{}'s performance has a large room to be improved for a higher estimation accuracy. Several approaches have been proposed, and one direction was to make \cmin{} better by replacing update mechanism~\cite{goyal2012sketch} or by changing counter representation~\cite{pitel2015cmlog,pitel2016cmtree}, but they have limitations in terms of feasibility in data plane or memory efficiency. Another direction was to cascade multiple sketches in a sequential manner so that either mouse or elephant flows should be filtered to separate elephants from mouse flows. Especially, those systems adopt a pyramid-shaped data structure that has more counters for small flows and less for elephant flows to reflect Zipfian distribution~\cite{YangZJCL17Pyramid,zhou2018cold,song2020FCM}. However, most of the schemes except FCM sketch are not applicable to the data plane because of the high complexity of its memory update algorithm.

To enhance \cmin{} sketch, therefore, we are facing two challenges: one is (a) how to make \cmin{}'s data structure accommodate more effectively Zipfian distribution, and the other is (b) how to make update and query work without delaying packet processing in the switch's data plane. \ours{} adopts a different combination of data structure and update algorithm, namely ``Split Counter'' and ``Minimum Update'' to resolve (a) and (b), respectively.
{\em Split Counter} design is to give more counters at a constant memory footprint. In contrast to \cmin{} that has the same number of counters ($w$) in $d$ layers where all counters are fixed-length (\eg 32 bits), wasting many bits for many mouse flows, \ours{} splits a counter into multiple counters of a smaller bit to allocate more counters of a small bit size to smaller flows of which number is much larger than that of large flows in Zipfian distribution. Taking advantage of this Pyramid-shaped data structure is not new. However, with our unique update strategy, called the cross-layer counting with a minimum update, or {\em Minimum Update} in short, a small size flow is decodable in every layer while keeping the number of counters to be updated minimum to give better estimation accuracy than \cmin{}. This contrasts with cascaded multi-sketch schemes adopting similar data structure such as ~\cite{YangZJCL17Pyramid,zhou2018cold,song2020FCM}, where a flow is decoded only at a single layer according to its size, resulting in an adverse effect to the accuracy. In short, ``Minimum Update'' can be regarded as an ASIC-friendly approximation (\ie pipeline design) of the famous conservative update, which is infeasible in the data plane (see section~\ref{sec:p4_impl} for details).

To sum up, being different from multi-sketch cascaded schemes, such as Pyramid sketch~\cite{YangZJCL17Pyramid}, Cold filter~\cite{zhou2018cold}, Elastic sketch~\cite{yang2018elastic}, and FCM sketch~\cite{song2020FCM}, our \ours{} sketch is a variant of \cmin{}, a single sketch with ``Split Counter'' data structure and ``Minimum Update'' algorithm (see Fig.~\ref{fig:filters} and Fig.~\ref{fig:CMCL}). \ours{}'s approach is shown to be superior in terms of estimation accuracy at a constant memory footprint and to be better in terms of data plane latency and packet processing throughput.

\BfPara{Contributions} Our contributions are as follows:

\begin{enumerate}
    \item Inherent memory waste of \cmin{} is investigated in detail, and memory inefficiency of FCM, the latest cascaded multi-sketch scheme, is analyzed comparatively to \ours{}. We present \ours{}, which has data structure reflecting Zipfian distribution and ASIC-implementable approximation of \cmcu{} sketch. \ours{} has better accuracy and higher throughput than state-of-the-art sketches while not requiring any additional data structures or hardware support, allowing an easy deployment. 
    \item Comprehensive analyses, including theoretical proof on the estimation bound of \ours{} and experimental analysis, are presented. Through comprehensive experiments using real network traces, we comparatively evaluate \ours{} alongside \cmin{}, \cmcu{}, Elastic sketch, and FCM sketch. \ours{} is shown to be superior to \cmin{}, \cmcu{}, Elastic sketch, and FCM, the state-of-the-art algorithms in terms of ARE (Average Relative Error), WMRE (Weighted Mean Relative Error), entropy, cardinality, heavy hitter detection, and heavy changer detection. 
    \item To show \ours{}'s feasibility for a high-speed switch, a prototype is implemented on a programmable switch (Tofino) in P4 language~\cite{tofino_wedge, p4_lan_spec}. Tailoring \ours{}'s parameters for the switch is necessary due to the programmable switch's hardware constraint. The hardware version's performance is comparatively analyzed to \ours{}'s software version. 
\end{enumerate}

\BfPara{Organization} The rest of the paper is organized as follows: we introduce our motivation, including constrains of ASIC-based switch, arguments for and against \cmin{} sketch, and the design choice of \ours{} sketch in section~\ref{sec:motiv}. In section~\ref{sec:main}, we introduce our \ours{} sketch with its analysis in section~\ref{sec:analysis}. We evaluate \ours{} sketch in section~\ref{sec:eval}. We implement \ours{} sketch with various applications in P4 switch and evaluate them in section~\ref{sec:p4_impl}. We review the literature in section~\ref{sec:relatedwork}, and conclude in section~\ref{sec:conclusion}.

\section{Motivating \ours}\label{sec:motiv}

\subsection{\cmin{}-based in-network traffic measurement system: Limitations} The recent trend of network research has been expanded to ASIC-based programmable switches using domain-specific languages (\ie P4~\cite{p4_lan_spec}), allowing advanced ideas to be deployed in the data plane. However, this flexibility is provided under various constraints (\ie memory size and computational complexity) to support the line-rate packet processing speed. 
As such, network traffic measurement algorithms remain untested for their feasibility in the ASIC-based programmable switch~\cite{huang2017sketchvisor,MoshrefYGV16Trumpet,zhou2018cold}. To date, several switch-based generic measurement system were proposed~\cite{li2016flowradar,liu2016one, yang2018elastic, song2020FCM}. However, these systems either provide an unsatisfying accuracy~\cite{li2016flowradar, liu2016one} or suffer from an over-reliance on advanced memory~\cite{yang2018elastic, song2020FCM}.

It is well known that \cmin{} provides a good estimation accuracy with a fixed and small amount of memory. More importantly, the generic measurement features allow \cmin{} to support various queries (\ie point, range, inner product, heavy hitter, and quantile). Especially, and thanks to its simplicity, \cmin{} can easily fit within the constrained switch environment, allowing systems like Elastic Sketch~\cite{yang2018elastic} and FCM Sketch~\cite{song2020FCM} to employ \cmin{} as an essential component for supporting generic measurement tasks (\ie cardinality, flow size distribution, entropy, heavy hitter, heavy changer). Even so, \cmin{} is not preferred as a standalone function in switch for network measurement tasks because \cmin{}'s simple data structure and algorithm designs have also become its main drawbacks, triggering memory waste and poor accuracy issues.  

\begin{figure*}[t]
    \centering
    \subfigure[Flow size distribution]{\includegraphics[width=0.25\textwidth]{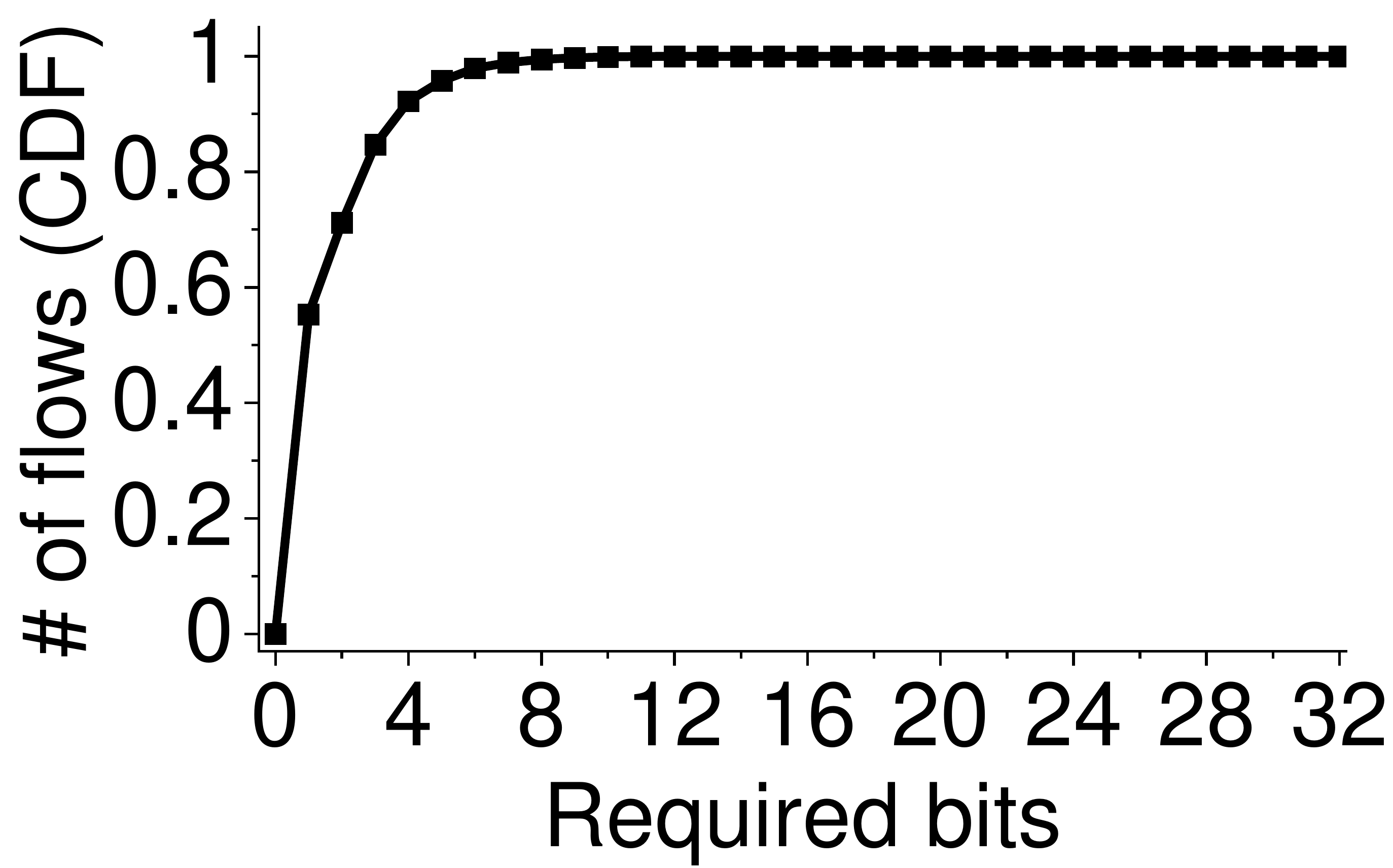}}\hspace{7mm}
    \subfigure[Distribution of wasted bits]{\includegraphics[width=0.25\textwidth]{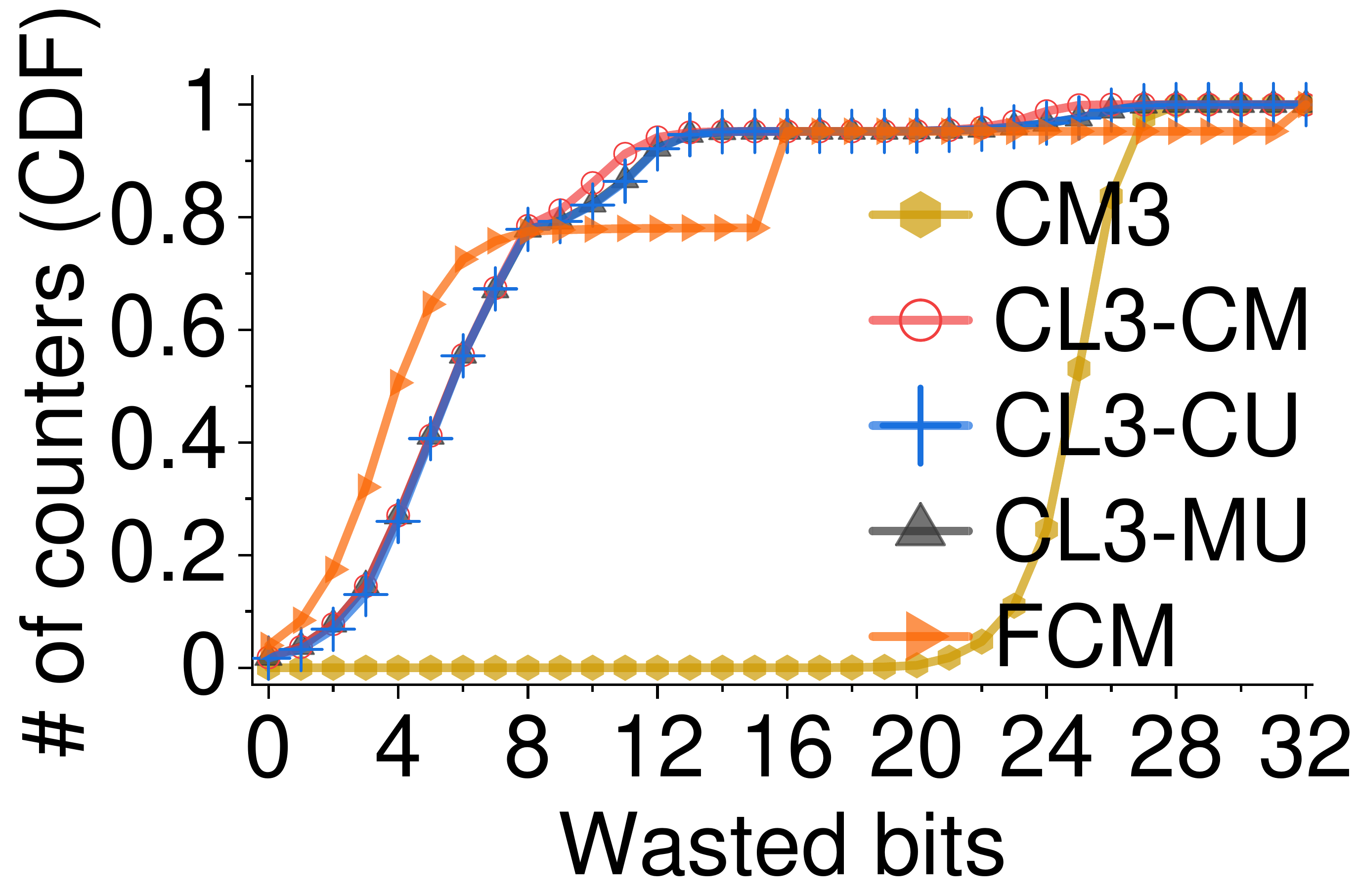}}\hspace{7mm}
    \subfigure[Accuracy varying flow size]{\includegraphics[width=0.25\textwidth]{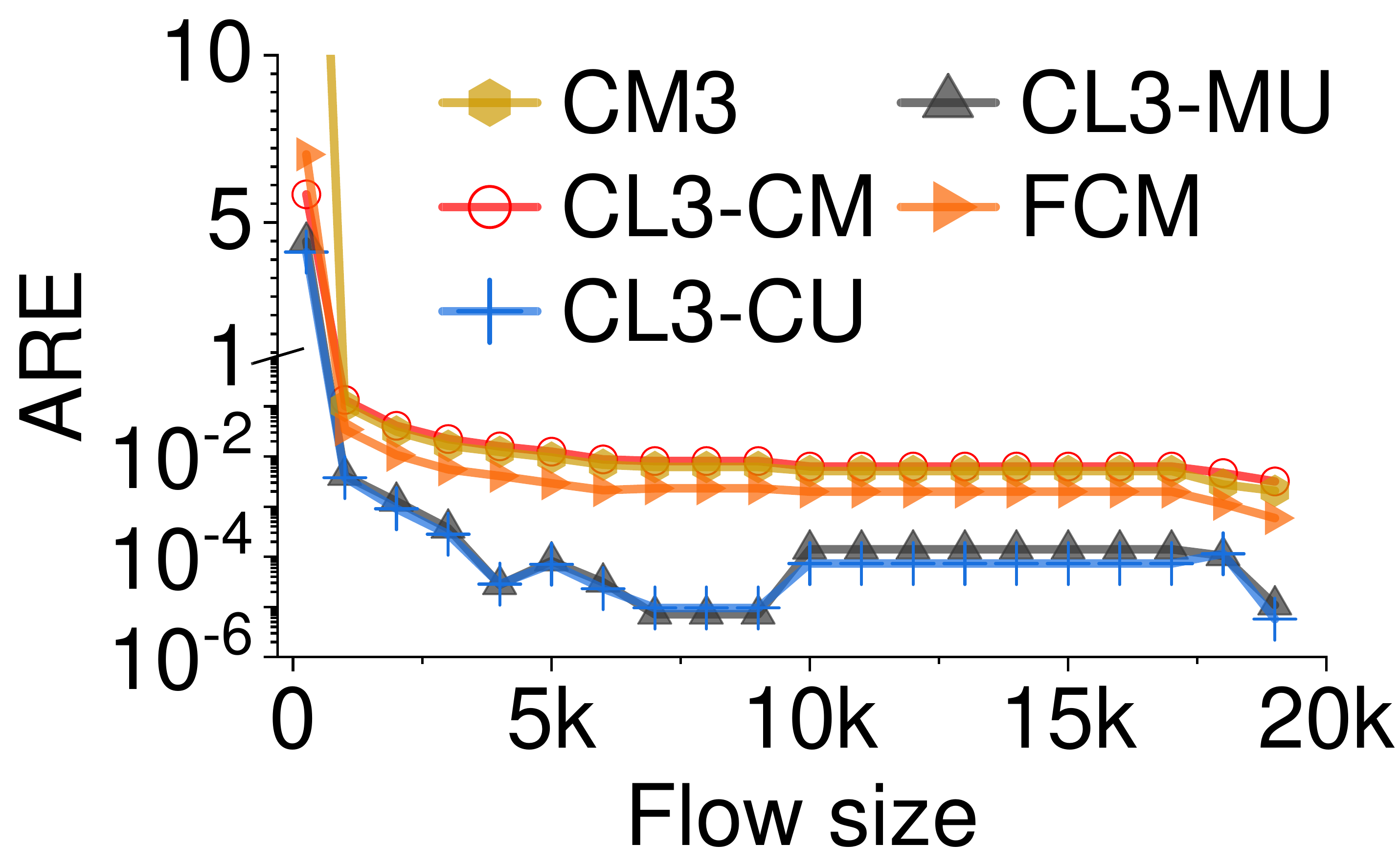}}
    \subfigure[\ours{}: distribution of wasted bits per layer]{\includegraphics[width=0.38\textwidth]{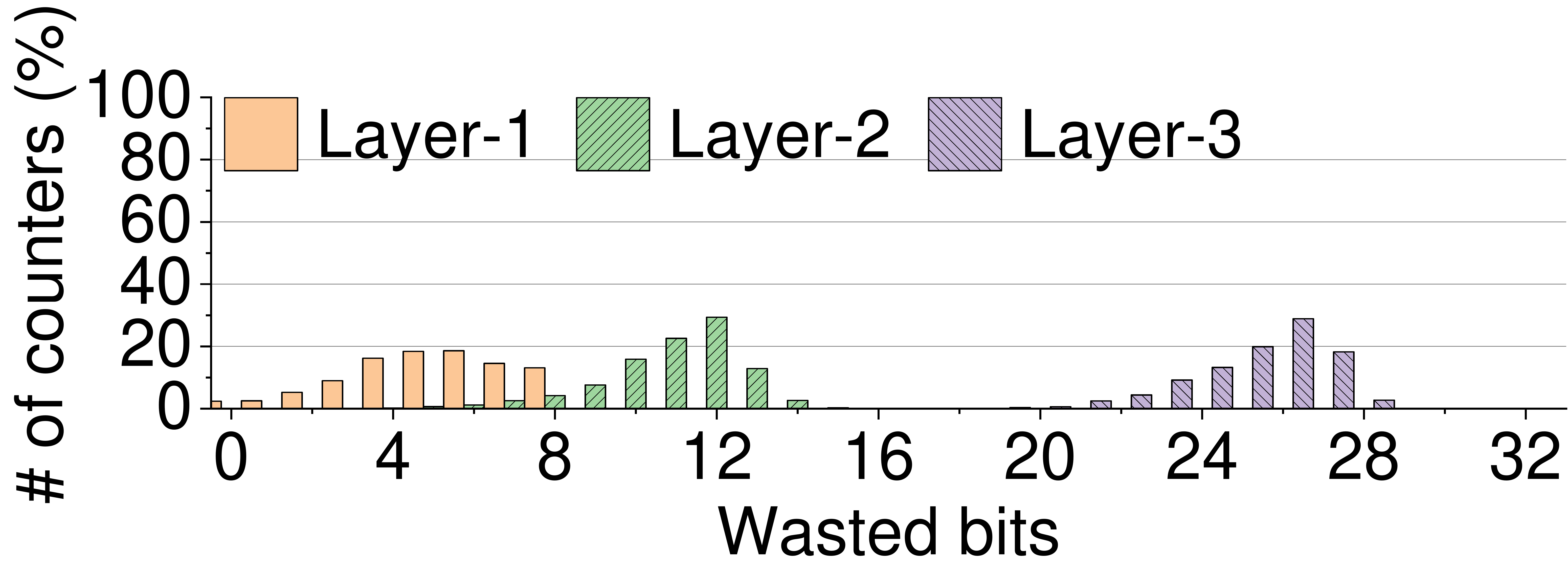}}\hspace{14mm}
    \subfigure[FCM sketch: distribution of wasted bits per layer]{\includegraphics[width=0.38\textwidth]{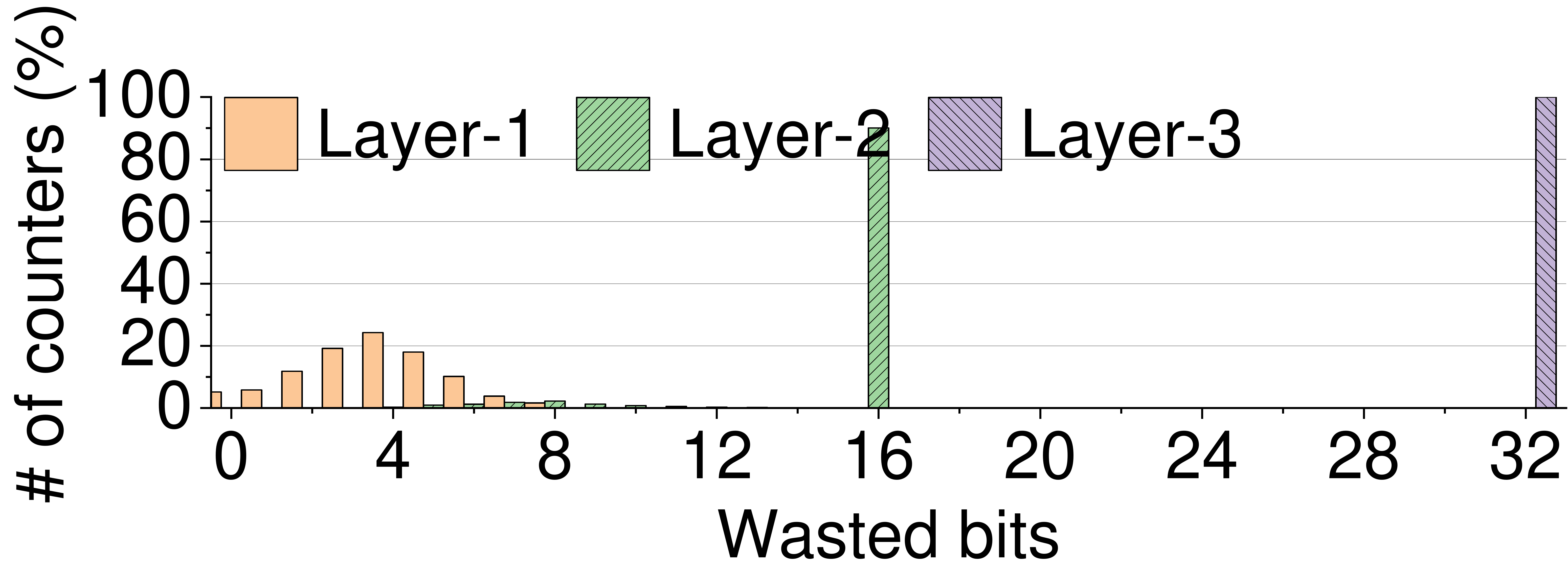}}
    
    \caption{Wasted bits and accuracy.
		     (a) shows the flow size distribution an example trace. (b) illustrates the distribution of wasted bits varying schemes. (c) demonstrates the relative error varying flow sizes and schemes. (d) and (e) compare the detailed distribution of wasted bits between our \ours{} and a state-of-the-art scheme, FCM sketch. The result suggests our scheme makes better use of the entire memory space for more accurate estimations than FCM, in which ARE reaches as high as 7.02 while CL3-MU has ARE of 4.40.
		}~\label{fig:mov}
\end{figure*}

As shown in Fig.~\ref{fig:CMCL}(a), the \cmin{}'s counter size is identical across all layers, which is well known to ignore the skewed distribution of modern network traffic (\ie massive mouse flows and sporadic elephant flows). For a 32-bit counter, counters occupied by mouse flow, which is the majority of network traffic, use only a few less significant counter bits. As a result, most of the counters' space remains unused, resulting in memory waste of sketch. Fig.~\ref{fig:mov}(a) shows the accumulated flow amount by varying the required bits for counting (\ie $log_2(f+1)$) with a five second CAIDA trace. As shown, about 92\% of the flows require less than 4 bits, and 99\% of the flows require less than 8 bits. Therefore, given a 32-bit counter-based \cmin{} with $d$=3, about 89.0\% of the counters waste more than 24 most significant bits in each counter, as the grey line shows in Fig.~\ref{fig:mov}(b), which is equivalent to 78.85\% of the entire memory. While the memory utilization increases as more traffic is recorded, the noise level will rise accordingly (\ie trade-off). Moreover, modern switches have a very limited memory, due to both speed and cost constraints, thus \cmin{} is unqualified as a data plane function.

\subsection{Multi-Sketch Cascaded Counters}
\begin{figure*}
    
    \begin{minipage}{0.65\textwidth}
    \subfigure[Cold Filter]{\includegraphics[height=4.5cm]{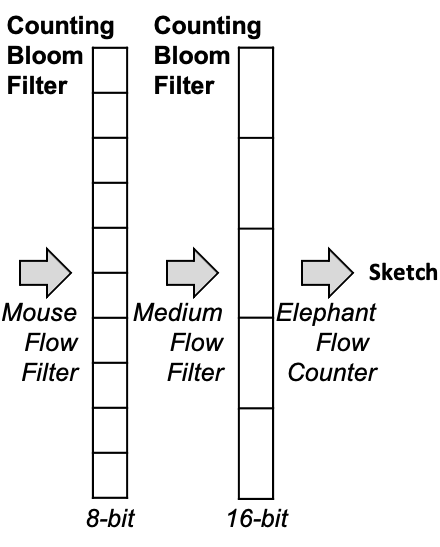}}\hspace{5mm}
    \subfigure[Elastic Sketch]{\includegraphics[height=4.5cm]{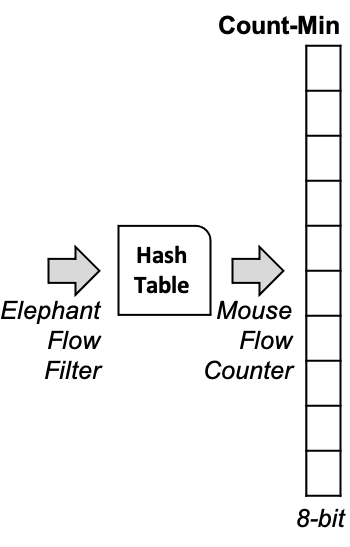}}\hspace{5mm}
    \subfigure[Pyramid and FCM]{\includegraphics[height=4.5cm]{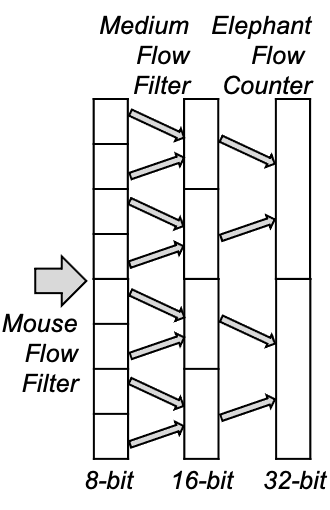}}
    \caption{
		    Cascaded multi-sketch approaches
		}~\label{fig:filters}
    \end{minipage}\hspace{-5mm}
    \begin{minipage}{0.35\textwidth}
    \centering
    \subfigure[Count-Min]{\includegraphics[height=4.5cm]{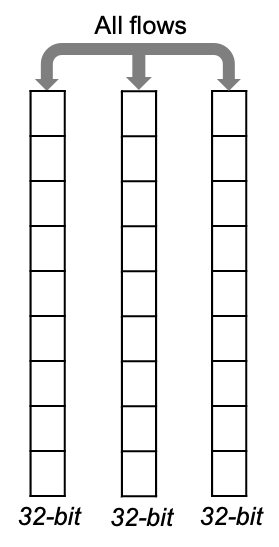}}\hspace{5mm}
    \subfigure[Count-Less]{\includegraphics[height=4.5cm]{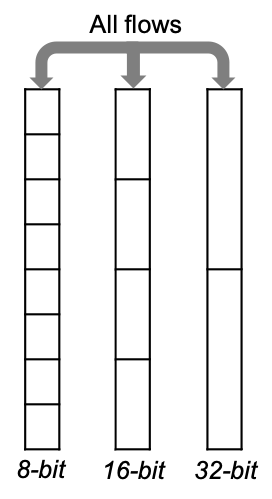}}
     \caption{
		    Cross-layer counting approaches
		}~\label{fig:CMCL}
    \end{minipage}
    
\end{figure*}

There have been several efforts to address inefficient data structures design for skewed data streams~\cite{yang2018elastic, YangZJCL17Pyramid,zhou2018cold,song2020FCM, estan2002new}. These approaches all fall in the same category, namely {\it multi-stage filtering}, which is a concept that was first introduced by Estan~\etal~\cite{estan2002new}. In these approaches, a number of sketches (\ie shared counter/bit arrays) are cascaded for a sequential flow filtering according to their size (\ie multi-sketch cascaded counters).  As shown in Fig.~\ref{fig:filters}(a), Cold Filter~\cite{zhou2018cold} concatenates two counting bloom filters with small counters (\eg 8-bit) for filtering out mouse and medium-sized flows sequentially, and finally stores elephants in an additional data structure, such as \cmcu{}, Space-Saving~\cite{metwally2005efficient}, FlowRadar~\cite{li2016flowradar}. On the contrary, Elastic Sketch~\cite{yang2018elastic} counts and filters out elephant flows first, and then records mouse flows using the \cmin{} sketch with 8-bit counters, as shown Fig.~\ref{fig:filters}(b). For systematic filtering, Pyramid Sketch~\cite{YangZJCL17Pyramid} presented a tree-based data structure that enlarges the size of each layer's counters as the decrements of the tree depth, as shown in Fig.~\ref{fig:filters}(c). Later, FCM sketch~\cite{song2020FCM} extended a similar idea with a multi-tree design and implemented it in a programmable switch (state-of-the-art). Although these approaches achieve better accuracy than \cmin{}, the biggest contributing factor to their higher accuracy is their data structure design, which allocates more smaller counters for mouse flows and fewer larger counters for medium-sized and elephant flows. While this data structure design can resolve the memory waste issue (Fig.~\ref{fig:mov}(b)), we stress that these solutions cannot fairly utilize the entire memory space due to the imbalanced task loads at each layer (\ie cascade counting), as shown in Fig.~\ref{fig:mov}(e). Similar to \cmin{}, even though the memory utilization of higher layers increases as more traffic arrive, the noise level of the lowest layer will rise accordingly, thus aggravate the flows overestimation. We note that Fig.~\ref{fig:mov}(e) demonstrates FCM sketch's per-layer distribution of wasted bits when sufficiently large traffic is recorded.

\subsection{Our Approach}
The eventual goal of our \ours{} is not only to take advantage of advanced data structures for memory efficiency but also to bound our sketch's estimation error under \cmin{}'s theory. Moreover, we aim to realize our algorithm in a switch's data plane pipeline. In the following, we introduce our observations and the design choices of our \ours{}. We note that different combinations of data structures and algorithms result in different behaviors of sketches.

\BfPara{(1) Split Counter} Our data structure design is similar to the Pyramid and FCM sketches, as shown in Fig.~\ref{fig:filters}(c), with the essential idea of splitting the counters into smaller pieces for reducing mouse flows' collision rate. As shown, the counter size of the lower layer is smaller than the corresponding ones in the upper layer. The size can be expressed as $b_1<b_2<\cdots<b_d$, where $b_i$ denotes the counter size at layer-$i$ and $d$ is the number of layers. By splitting the counters, the total number of counters ($w\cdot d$) increases significantly as $w'_1+w'_2+w'_3+\dots+w'_d \gg w\cdot d$, thus reducing the hash collision probability of flows at each layer.

\BfPara{(2) Cross-layer Update} For update, the multi-sketch cascaded approaches encode each flow at a single layer and cascade the counter value to the next layer when the current layer's counter is overflowed. To further reduce the hash collision and increase the flow survival rate, we suggest enabling cross-layer counter update like \cmin{}, which can achieve a tighter error bound.

{\it \cmin{}'s Update Algorithm (baseline):} To explore the design space, we first use the split counters (\ie pyramid counters) with \cmin{}'s update algorithm, which increases a flow's counters at all layers simultaneously (hereafter \clcm{}). As shown in Fig.~\ref{fig:mov}(c), \clcm{} is more accurate than \cmin{} for mouse flow estimation  (\ie $< 2^8-1$) but has a higher relative error for medium-sized and elephant flows. As expected, the mouse flows benefit from the large number of small counters. Moreover, this \clcm{} reduces the memory waste rate from \cmin{}'s 78.85\% to around 67.64\%, as shown in Fig.~\ref{fig:mov}(b).
[{\bf Observation 1.} We observed that the cross-layer update algorithm allows mouse flows to be hosted by idle counters at all layers for having more chances to survive.]
As a result, \clcm{} achieves better mouse flows estimation than the cascaded counter, FCM sketch, which records mouse flows only at the first layer, as shown in Fig.~\ref{fig:mov}(e).
Nevertheless, \cmin{}'s poorer accuracy for larger flows exposes its drawback due to the indiscriminate cross-layer update of counters. 
In other words, even though mouse flows survive well, these flows flood counters at all layers, increasing the noise level of larger flows. To this end, the subsequent challenge is {\it how to prevent mouse flows from flooding other flows in the cross-layer update mechanism?}

{\it Conservative Update (challenge):} [{\bf Observation 2.} Inspired by the conservative update of \cmin{}, which achieves a better accuracy than the standard \cmin{}, we observed a similar result when applying the conservative update with the split counters (hereafter, \clcu{}), as shown in Fig.~\ref{fig:mov}(c)]. While we consider this approach to be {\it optimal}, we are not able to realize the conservative update in a switch's data plane due to its pipeline design. Specifically, the conservative update algorithm must first determine the minimum counter(s) across all layers and revisit the counter for an update operation. However, the switch's data plane works as a pipeline, which means a double-access of a memory region is not allowed after the memory processing stage~\cite{zhang2020poseidon, Barefoot_Doc}. Therefore, the subsequent challenge becomes {\it how to update only minimum counter value(s) across all layers without memory double-access for accommodating a switch's data plane?}

{\it Minimum Update (our solution):} 
Our update algorithm, called {\em Minimum Update} or simply {\em CL-MU}, is designed to update as fewer counters as possible, and only the minimal counter value(s) of a flow across all layers without revisiting counters. 
We note that both concepts are crucial for \ours{}. With {\it Split Counter}, \ours{} maximizes the total number of counters and adapts to a skewed data stream.  With {\it Minimum Update}, \ours{} reduces the memory waste, as shown in Fig.~\ref{fig:mov}(b), thanks to the cross-layer update mechanism. Moreover, unlike the cascaded approaches, \ours{} can leverage idle counters at all layers to host more mouse flows, as shown in Fig.~\ref{fig:mov}(d) and Fig.~\ref{fig:mov}(e). Moreover, \ours{} gives a higher priority to elephant flows at the highest layer, thereby preventing larger flows from being flooded by mouse flows due to our minimum update strategy. As a result, given the same amount of memory, \ours{} outperforms the standard \cmin{} and the state-of-the-art FCM sketch, and achieves comparable accuracy with the optimal solution (\ie \clcu{} in all flow sizes, as shown in Fig.~\ref{fig:mov}(c). Finally, \ours{}'s operation does not require a double-access of counters like \clcu{}, therefore it is deployable in a switch's data plane and requires less resource than state-of-the-art approaches (see section~\ref{sec:p4_impl}).

\section{The Count-Less Sketch}\label{sec:main}
In this section, we describe the data structure and algorithm of our per-flow counting sketch in more details.

\subsection{Data Structure}
As shown in Fig.~\ref{fig:CMCL}(b), \ours{}'s consists of $d$ counter arrays. Unlike \cmin{}, \ours{} differentiates both the counter size and array size at each layer to (1) reduce the bit waste of counters and (2) adapt to the skewed flow distribution of modern network traffic. For the counter size, the highest layer of \ours{} uses 32-bit counters to support a sufficient counting range. Then, we make the size of counters half for each layer while going through the lowest layer. 
We note that a similar data structure design appears in the Pyramid sketch~\cite{YangZJCL17Pyramid} and FCM sketch~\cite{song2020FCM}. However, \ours{} uses an update strategy that  behaves  differently compared to the previous sketches. In particular, 
\ours{} assigns a different number of counters for each layer to address the skewness of traffic. To do so, we used the factor $r$ to let a lower layer array possess $r$ times more counters than its upper layer. That is, $w_1 = rw_2 = \ldots = r^{d-2} w_{d-1} = r^{d-1}w_d$, where $w_i$ is the number of counters of the $i$-th layer.


\begin{figure*}[t]
    \centering
    \subfigure[Cross-layer update]{\includegraphics[height=3.8cm]{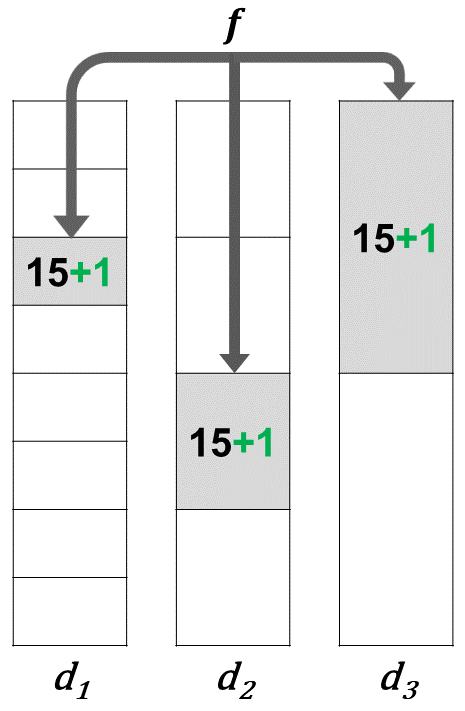}}\hspace{10mm}
    \subfigure[Minimum update: case 1. ]{\includegraphics[height=3.8cm]{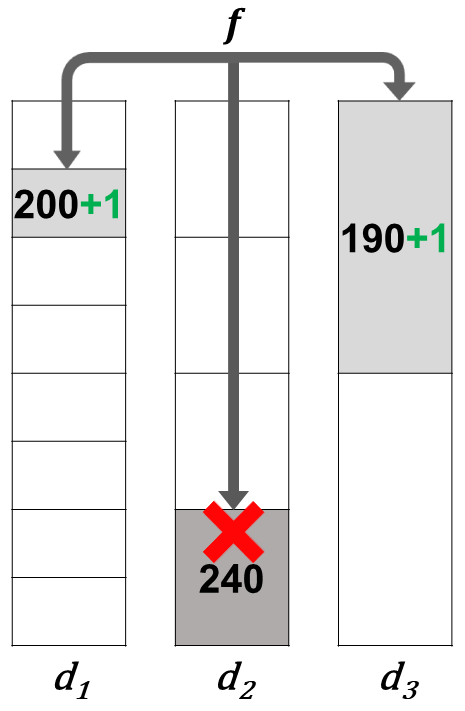}}\hspace{10mm}
   \subfigure[Minimum update: case 2.]{\includegraphics[height=3.8cm]{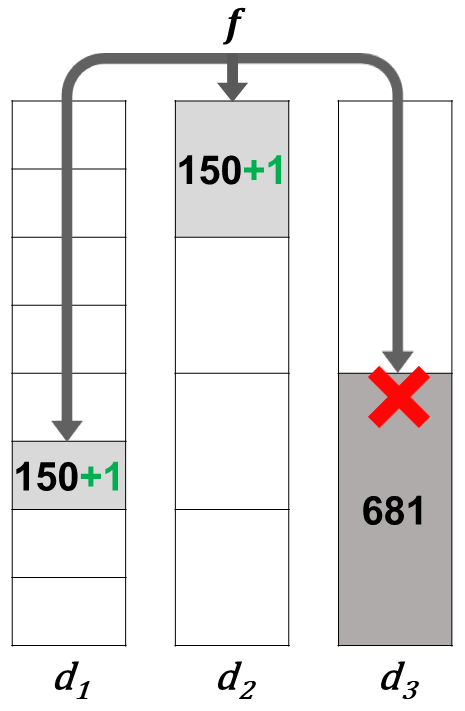}}\hspace{10mm}
    \subfigure[Worst case]{\includegraphics[height=3.8cm]{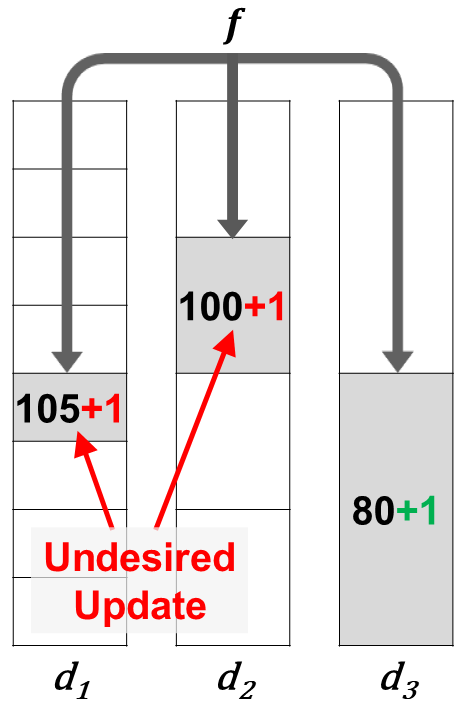}}
    \caption{
		     Examples of update operations. (a) shows that \ours{} allows mouse flows to survive at any layer with idle counters. (b) and (c) depict that minimum update gives higher priority for elephant flows. (d) demonstrates the worst case update scenario of minimum update due to the unawareness of the global minimum. 
		}~\label{fig:operations}
\end{figure*}

\setlength{\textfloatsep}{6pt}
\begin{algorithm}[t]
{\footnotesize
    \textbf{Inputs}: Layer $d$, width of each layer $w_l$.\\
        $val_{min} = $\texttt{INT\_MAX}\;
        \For{l=1 to d}{
            $idx_{l} = hash_l(pkt)\ mod\ w_{l}$\;
            \If{$C_{l}[idx_{l}]\not=  overflow$}{
                \If{$C_{l}[idx_{l}] \sy{<} val_{min}$}{
                $val_{min} = ++C_{l}[idx_{l}]$\; 
                }
            }
        }
    \textbf{Return} $val_{min}$; /* Decoding on the fly*/\\
}
\caption{Encoding and Decoding}
\label{alg:hw_encoding}
\end{algorithm}

\subsection{Minimum Update}

\BfPara{(1) Encoding and Decoding} Algorithm~\ref{alg:hw_encoding} shows the encoding and decoding processes of \ours{} with {\em Minimum Update} (CL-MU). As shown in this figure, CL-MU's encoding and decoding are one combined operation. For every arriving packet, an estimation for the corresponding flow can be obtained while encoding, enabling the switch to detect anomalies on the fly (line 10). \ours{} updates one counter at each layer following the lowest-to-highest order (line 2). Moreover, instead of indiscriminately updating all counters like in \cmin{}, \ours{} maintains a variable for storing the minimal value ($val_{min}$) until the iterated layer (lines 1 and 6), and then updates the layer only if the counter ($C_l[idx_l]$) is smaller than $val_{min}$ (lines 5-6). The counter update event at a layer will be skipped when the layer's counter exceeds its counting capacity (line 4). As a result, \ours{} tends to update only the minimal counter value(s) of a flow across all layers, thereby a minimum number of counters will be updated with the worst case of all layer updates.

\BfPara{(2) Intuitions} We note that this {\em Minimum Update} strategy is unique to the {\em Split Counter} data structure. Since \ours{} maintains a temporal minimal value whenever updating a flow's counter at each layer and only updates the smallest counter upon discovery, \ours{} prevents an arriving flow from flooding counters simultaneously at all layers. 
Moreover, \ours{} can mitigate the flow counter collision in both horizontal and vertical manners. 
{\bf (1)} For mouse flows, the massive number of counters at the lower layers sufficiently reduce the collision probability, which results in mouse flows likely surviving within a lower layer (\ie intra-layer counter sharing mitigation). More importantly, our cross-layer update strategy allows mouse flows to leverage idle counters at all layers for surviving without interfering with the larger counters (\ie inter-layer counter sharing mitigation), as shown in the example in Fig.~\ref{fig:operations}(a) and (b).
{\bf (2)} On the other hand, elephant flows will survive at the highest layer due to the sufficiently large counters. 
More importantly, the elephant flows are given a higher priority, thereby will not be updated (or contaminated) by mouse flows due to the {\em Minimum Update}, as shown in the example in Fig.~\ref{fig:operations}(c). As a result, \ours{} eventually achieves a high flow survival rate with a fixed memory.

\BfPara{(3) Costs for fitting a switch's pipeline } The conservative update algorithm requires awareness of the global minimum value across all layers for a flow before counter update. However, this process triggers a double-access of the smallest counter and is unacceptable for the pipeline-based switches (\ie see section~\ref{sec:p4_impl} for the memory double-access constraint). To address this issue, \ours{} approximates the global minimum value with the temporal minimum value upon the iterated layers. As shown in Fig.~\ref{fig:operations}(d), this approximation results in some undesired updates at the beginning layer (\ie unawareness of the global minimum). However, we note that the approximation-caused error is negligible (see section~\ref{sec:eval}), making the minimum update algorithm feasible in the data plane.

\section{Theoretical Analysis}\label{sec:analysis}
In this section, we provide a theoretical analysis of \ours{}. In particular, the effects of the combination of {\em Split Counter} and {\em Minimum Update} is presented in terms of theoretical error bound and probability.
Since \ours{} follows the same update principle of the cross-layer update, we can take advantage of \cmin{}'s theory to calculate the error and the probability.
From \cite{cormode2005improved}, \cmin{} with an array of counters of width $w$ and depth $d$ has an error bound for a query $f$ for each row as follows~\cite{cormode2005improved}.


\begin{theorem}
Consider \ours{} sketch with Minimum Update having $d=\ln{1/\delta}$ layers for recording of flow size, where the $j$-th layer has $w_j=r^{d-j}w_d$ counters for $j=1,2, \ldots, d$. $w_d=e/\epsilon_d$ is the number of counters in the last ($d$-th) layer. Each counter in the $j$-th layer is $2^{j+1}$-bit long. 
\ours{} estimates $a_i$, the size of a flow $i$ by no larger than 
\begin{equation}
\hat{a_i} < a_i + 0.52\epsilon_d\lVert a^1 \rVert_{1_1'}
\end{equation}
with probability at least
\begin{equation}
Pr[\exists j, X_{i,j} < \epsilon_d \lVert a^d \rVert_{1_1'}] > 1-\delta^*,
\end{equation}
where $\delta^* = \delta*\prod_{j=1}^d r^{-d(d-j)}$.
\end{theorem}

{\noindent\em Proof:} Let $a^j_i$ be the actual flow size of the flow $i$ at layer $j$.
\ours{}'s estimation of $a_i$ is $\hat{a}_{i} = \textrm{min}_j(count[j, h_j(i)]),$ where $i\in[1, w_j=e/\epsilon_j], j\in[1, d=\ln{1/\delta}]$. 

Because of the additive errors by other flows, $count[j, h_j(i)] = a^j_i+X_{i,j}$, where $X_{i,j}$ is a random variable representing the added error to $a^j_i$ in the $j$-th layer. We define $a^j$ as the sum of all flows in the $j$-th layer:
\begin{equation}
    a^j \defeq \sum_{k=1}^n a^j_k
\end{equation}

The counters in the $j$-the layer of \ours{} has $2^{j+1}$ bits, and so the counting limit in each layer is defined by the number of bits. We denote the limits as $T_1, T_2, \ldots, T_{d-1}$ for the $d$ layers, respectively. Because of the limit, the total number of packets inserted in each layer  
\begin{equation}
\lVert a^j\rVert_{1_j'} \defeq \sum_{k=1,a_k^j \geq T_{j-1}}^n a_k^j 
\end{equation}
while $L_1$ norm of $a^j_i$ is defined by $\lVert a^j \rVert_1 = \sum_{k=1}^{n} a^j_k$.

For analysis of the error term, we introduce an indicator variable $I_{i,k,j}$ that indicates if there is a hash collision of the index $i$ and $k$ ($k\ne i$) with the hash function $h_j$. That is,

\begin{equation}
\begin{split}
I_{i,k,j} = 1 \leftrightarrow h_j(i) = h_j(k) \textrm{ for } i\ne k \textrm{ and } I_{i,k,j} \\ = 0 \textrm{ otherwise}
\end{split}
\end{equation}

The indicator $I_{i,k,j}$ being a binary random variable, the expectation of $I_{i,k,j}$ is the probability that $I_{i,k,j}=1$:
\begin{equation}\label{eq:ei}
\begin{split}
E[I_{i,k,j}] = Pr[h_j(i) = h_j(k)] \\ = 1/range(h_j) = 1/w_j = \epsilon_j/e
\end{split}
\end{equation}
assuming that $h_j()$ is chosen from a family of universal hash functions. Using the indicator, we can express the error term $X_{i,j}$ as follows:

\begin{equation}
X_{i,j} = \sum_{k=1}^n I_{i,k,j}*a^j_k = \sum_{k\vert h_j(i)=h_j(k), a^j_k \geq T_{j-1}}a^j_k
\end{equation}

Because $I_{i,k,j}$ and $a_k$ are independent, the expectation of the error is
\begin{equation}
\begin{split}
E[X_{i,j}] = E[\sum_{k=1, a^j_k \geq T_{j-1}}^n I_{i,k,j}*a^j_k] \\  = \sum_{k=1, a^j_k \geq T_{j-1}}^n a^j_k* E[I_{i,k,j}],
\end{split}
\end{equation}

Owing to equation~(\ref{eq:ei}), 
\begin{equation}\label{eq:ex}
E[X_{i,j}]= \sum_{k=1, a^j_k \geq T_{j-1}}^n a^j_k* E[I_{i,k,j}] = \lVert a^j \rVert_{1_j'}* \epsilon_j/e
\end{equation}

By Markov inequality and equation~(\ref{eq:ex}), the probability that the estimation error at a layer $j$ is greater than $\epsilon_j\lVert a^j \rVert_{1_j'}$ is as follows:

\begin{equation}\label{eq:er}
\begin{split}
    Pr[X_{i,j} > \epsilon_d \lVert a^d \rVert_{1_j'}] \leq Pr[X_{i,j} \geq \epsilon_d \lVert a^d \rVert_{1_j'}] \\ \leq \frac{\epsilon_j\lVert a^j \rVert_{1_j'}}{e\epsilon_d\lVert a^d \rVert_{1_j'}} < \frac{\epsilon_j\lVert a^d \rVert_{1_j'}}{e\epsilon_d\lVert a^d \rVert_{1_j'}} = \frac{\epsilon_j}{e\epsilon_d}
\end{split}
\end{equation}

In \ours{}, the number of layers used by a flow varies depending on its size by the {\em cross-layer update} strategy. 
Among $d$ layers, a flow of which size ranges from 1 to $T_1$ uses all the $d$ layers, while one from $T_1$ to $T_2$ uses 
$d-1$ layers because the lowest layer is saturated quickly and it does not update it anymore. Similarly, a flow of $T_{d-1}$ or above uses only one layer of \ours{}. The probability that the error term is bounded is thus all different according to a flow size. We can divide the probability and the error size estimation into $d$ cases with corresponding thresholds ($T_1, T_2, \ldots, T_{d-1}$, but we are going to show only three cases: the first case (the smallest flow group ($<T_1$), the $j$-th smallest flow group (between $T_{j-1}$ and $T_j$), and the largest flow group ($\geq T_{d-1}$). 

{\em Case 1)} We consider a flow $i$ in the smallest group that is less than $T_1$. In this case, a flow $i$ does not saturate even the lowest layer and has $d$ number of layers to record its size. By the pairwise independent hash assumption and equation~(\ref{eq:er}), the probability that for all layers the error is greater than $\epsilon_j \lVert a^j \rVert_{1_j'}$ is as follows:
\begin{equation}
Pr[\forall j, X_{i,j} > \epsilon_j \lVert a^j \rVert_{1_j'}] < \prod_{j=1}^{d} \frac{E[X_{i,j}]}{\epsilon_j\lVert a^j \rVert_{1_j'}} =  \frac{1}{e^d} = \delta
\end{equation}
    
The probability is bounded by $\delta$ even with the skewed data structure of \ours{}, but the error bound for this flow size group is different in each layer. 
\ours{} as well as \cmin{} estimates $a_i$ by the minimum among the counters. 
That is, the error of the $j$-th layer is bounded as follows:
\begin{equation}
\hat{a_i} < a_i + \epsilon_j\lVert a^j \rVert_{1_j'}
\end{equation}
with probability at least
\begin{equation}
Pr[\exists j, X_{i,j} < \epsilon_j \lVert a^j \rVert_{1_j'}] > 1-\delta,
\end{equation}

{\em Case 2)} For a flow of size ranging from $T_{j-1}$ to $T_j$, a flow $i$ in this group already saturates the lower $j-1$ layers, and it is allowed to use $d-j+1$ layers. Owing to \ours{}'s MU update policy, the flow group overrides any flow in smaller flow groups, while larger flows saturate and do not use this layer anymore. From the view of a flow in this group, this case is equivalent to having this layer for its exclusive use, while sharing counters with smaller flows that are not collided. For a flow $i$, thus, the number of packets of interest in this layer is $\lVert a^j\rVert_{1_j'}$.

The $j$-th and $j+1, \ldots, d$ layers are used for recording the flow size of this group, thus the probability that the error in all of $d-j+1$ layers goes beyond the bound of equation~(\ref{eq:er}) is:
\begin{equation}
\begin{split}
Pr[\forall s\in [j,\ldots,d], X_{i,s} > \epsilon_s \lVert a^s \rVert_{1_s'}] \\ < \prod_{s=j}^{d} \frac{E[X_{i,s}]}{\epsilon_s\lVert a^s \rVert_{1_s'}} = \frac{1}{e^{d-j+1}}. 
\end{split}
\end{equation}

\ours{} estimates $a_i$ by no larger than $\epsilon_j\lVert a^j \rVert_{1_j'}$ with a probability greater than $1-1/e^{d-j+1}$.
\begin{equation}
\hat{a_i} < a_i + \epsilon_j\lVert a^j \rVert_{1_j'},
\end{equation}
with probability of at least
\begin{equation}
Pr[\exists j, X_{i,j} < \epsilon_j \lVert a^j \rVert_{1_j'}] > 1-1/e^{d-j+1}. 
\end{equation}

{\em Case 3)} For a flow of size ranging from $T_{d-1}$ to $T_d$, a flow $i$ is in the largest group, which saturates every layer except the last layer, thus has only one layer to record its size. Owing to \ours{}'s MU update policy, the flow group overrides any flow in smaller flow groups. From the view of a flow in this group, this case is also equivalent to having this layer for its exclusive use. However, we note that unlike the multi-stage sketches, smaller flows without hash collision with the large flows still can record their sizes in this layer. Similarly, we have:
\begin{equation}
Pr[X_{i,d} > \epsilon_d \lVert a^d \rVert_{1_d'}] < \frac{E[X_{i,d}]}{\epsilon_d\lVert a^d \rVert_{1_d'}} =  \frac{1}{e}
\end{equation}
\ours{} estimates $a_i$ by no larger than $\epsilon_1 \lVert a^d \rVert_{1_d'}$ with a probability greater than $1-1/e$. 

\begin{equation}
\hat{a_i} < a_i + \epsilon_d\lVert a^d \rVert_{1_d'}
\end{equation}
with probability at least
\begin{equation}
Pr[\exists j, X_{i,d} < \epsilon_d \lVert a^d \rVert_{1_d'}] < 1-\frac{1}{e}
\end{equation}

Because each layer has a different error bound and corresponding probability, we are going to provide an average of those error bounds over the $d$ layers. Moreover, we  conservatively provide a probability of the estimation that \ours{}'s error is less than or equal to $\epsilon_d \lVert a^d \rVert_{1_d'}$. We note that if we set $w_d = w_{CM}$, this bound is equal to \cmin{}'s bound.  

Assuming that the number of packets follows a Zipf distribution with a Zipf parameter of 1, the average error bound of \ours{} over all the $d$ layer is calculated as follows:

\begin{equation}
\begin{split}
&E(\epsilon_j*\lVert a^j \rVert_{1_j'}) = E(\epsilon_j)E(\lVert a^j \rVert_{1_j'})\\
&= E(\epsilon_j)\sum_{j=1}^d \lVert a^j \rVert_{1_j'}/{d} = E(\epsilon_j) \sum_{j=1}^d \sum_{a_k^j \geq T_{j-1}} a_k^j/d\\
&= E(\epsilon_j)(d a^1 + (d-1) a^2 + \ldots + 1 a^d)/d \\
&= E(\epsilon_j)\lVert a^1 \rVert_{1_1'}(d/1 + (d-1)/2 + \ldots + 1/d)/d \\ 
&= (\sum_{j=1}^d \epsilon_j/d) \lVert a^1 \rVert_{1_1'}(d/1 + (d-1)/2 + \ldots + 1/d)/d \\ 
&= (\sum_{j=1}^d e/(w_jd)) \lVert a^1 \rVert_{1_1'}(d/1 + (d-1)/2 + \ldots + 1/d)/d \\ 
&= (\sum_{j=1}^d e/w_j)\lVert a^1 \rVert_{1_1'}(d/1 + (d-1)/2 + \ldots + 1/d)/d^2 \\ 
&= (\sum_{j=1}^d e/(r^{d-j}w_d))\lVert a^1 \rVert_{1_1'}(d/1 + (d-1)/2 + \ldots + 1/d)/d^2 \\ 
&= (e/w_d)(1/r^d)(\sum_{j=1}^d r^j)\lVert a^1 \rVert_{1_1'}(d/1 + (d-1)/2 + \ldots + 1/d)/d^2 \\ 
&= (\frac{r(r^d-1)}{r^d(r-1)})\epsilon_d\lVert a^1 \rVert_{1_1'}(d/1 + (d-1)/2 + \ldots + 1/d)/d^2
\end{split}
\end{equation}

For  $d=3, 4, \ldots, 6$, $(d/1 + (d-1)/2 + \ldots + 1/d)/d^2 < 0.48$. 

For  $r=4, {r(r^d-1)}/{r^d(r-1)} \approx 1.3$.  Therefore,
\begin{equation}
    E(\epsilon_j*\lVert a^j \rVert_{1_j'}) < 0.52\epsilon_d\lVert a^1 \rVert_{1_1'}
\end{equation}


\ours{}'s estimation error goes beyond $\epsilon_d\lVert a^d \rVert_{1}$ when the corresponding counters in all the $d$ layers of \ours{} are greater than $\epsilon_d\lVert a^d \rVert_{1}$. Thus, taking advantage of equation~(\ref{eq:er}) and the independent hash assumption among layers, the probability is calculated as follows: 

\begin{equation}
\begin{split}
Pr[\forall j, X_{i,j} > \epsilon_d \lVert a^d \rVert_1] = \prod_{j=1}^{d} \frac{\epsilon_{j}/e\lVert a^j \rVert_1}{\epsilon_{d}\lVert a^d \rVert_1} \\ = \prod_{j=1}^{d} \frac{\epsilon_{d}/r^{d-j}\lVert a^j \rVert_1}{e*\epsilon_{d}\lVert a^d \rVert_1}
< \prod_{j=1}^{d} \frac{\epsilon_{d}/r^{d-j}\lVert a^d \rVert_1}{e*\epsilon_{d}\lVert a^d \rVert_1}\\
= \prod_{j=1}^{d} \frac{1}{e*r^{d-j}} = e^{-d} * \prod_{j=1}^d r^{-d(d-j)} = \delta * \alpha,
\end{split}
\end{equation}

where $\alpha = \prod_{j=1}^d r^{-d(d-j)}$ is less than 1 (\eg $\alpha \approx 3.8*10^{-6}$ for $r=4, d=3$). 
Therefore,
\begin{equation}
Pr[\exists j, X_{i,j} < \epsilon_d \lVert a^d \rVert_{1_1'}] > 1-\delta^*,
\end{equation}
where $\delta^* = \delta*\prod_{j=1}^d r^{-d(d-j)}$. $\QEDA$\\

We note that the expectation of the error term of \ours{} is less than that of \cmin{} while providing better probability as shown above, when we set $w_d = w_{CM}$.

\section{Evaluating \ours{}}\label{sec:eval}
Our evaluations are threefold. First, we analyze the effect of three important strategies (Split Counter, Cross-layer Update, and Minimum Update) that \ours{} is composed of. The analysis is comparatively done with various solutions, including (1) standard (\cmin{}), (2) optimal (\clcu{}), and (3) state-of-the-art  (FCM sketch~\cite{song2020FCM}). Second, we vary \ours{}'s parameters to find the appropriate settings. Finally, we perform six network traffic measurement tasks with \ours{} and compare the performance with the other sketches such as Elastic sketch~\cite{yang2018elastic} and FCM sketch~\cite{song2020FCM}.

\subsection{Dataset and Parameters} 
In our experiments, we used a one-hour real-world trace from CAIDA~\cite{CAIDA}. We divided the trace into 720 sub-traces with a five second interval and used the first 32 sub-traces considering each sub-trace to belong to an epoch. Each sub-trace contains 2.5 M to 2.7 M packets with 226 K to 244 K distinct 5-tuple flows (\ie SrcIP, DstIP, SrcPort, DstPort, and protocol). 
We also used one-minute traces of 31.27 M packets with 1.88 M flows to show \ours{}'s characteristics under an extreme case. 
To show the memory impact, we varied the used memory from 0.2 MB to 1 MB for all schemes. For \ours{}, we set $d = 3$ and $d = 4$ with $r = 4$ (see section~\ref{sec:param_config} for details about parameter configuration). For fairness, the number of layers ($d=3$) is identical for all \cmin{}-based schemes. Moreover, Elastic and FCM sketches' configurations followed the configurations in the original works. Namely, for Elastic, we assign 150 KB for memory for the hash table. For FCM, we set $d=2$ and $k=4$ for a fair comparison. 

\subsection{Evaluation Metrics} 
In this paper, we use six metrics to evaluate \ours{}, as follows:

\BfPara{(1) ARE (Average Relative Error)} ARE is the averaged relative errors of flows, $\frac{1}{n}\sum_{i=1}^{n}{f_{i}-\hat{f_{i}}}/{f_{i}}$, where $n$ is the number of flows and $f_{i}$ and $\hat{f_{i}}$ are the actual and estimated flow sizes, respectively. ARE is used to evaluate the accuracy of the flow size estimation.

\BfPara{(2) Flow Survival Rate} We define the flow survival rate as the fraction of flows that are below a certain relative error after decoding. Considering the different noise impacts for different sized flows, we break down the flow sizes in three ranges, namely 1$\sim$254, 255$\sim$65,534, and 65,535$\sim$, and consider a flow as survived if the estimated relative error is below 0.1, 0.05, and 0.01, respectively. 

\BfPara{(3) RE (Relative Error)}
$\left | 1-\frac{estimated}{actual} \right |$, with the {\em actual} and {\em estimated} values, respectively. We use RE to evaluate the accuracy of the cardinality and entropy estimations.

\BfPara{(4) Weighted Mean Relative Error (WMRE)}
$\frac{\sum_{i=1}^{z}\left | n_{i}-\hat{n_{i}} \right |}{\sum_{i=1}^{z}(\frac{n_{i}+\hat{n_{i}}}{2})}$, where $z$ is the maximum flow size,  $n_{i}$ and $\hat{n_{i}}$ are the actual and estimated numbers of flow size $i$,  respectively~\cite{huang2017sketchvisor, kumar2004data}. We use WMRE to evaluate the accuracy of the flow size distribution.

\BfPara{(5) F1 Score}
$2 \times \frac{precision \times recall}{precision + recall}$, where the precision refers to the ratio of the true instances reported and recall refers to the ratio of the reported true instances. We use the F1 score to evaluate the accuracy of the heavy hitter and heavy changer.

\BfPara{(6) Throughput}
Million packets per second (Mpps) indicates the packet processing capacity.

\begin{figure*}[t]
    \centering
    \subfigure[Number of Counters]{\includegraphics[width=0.23\textwidth]{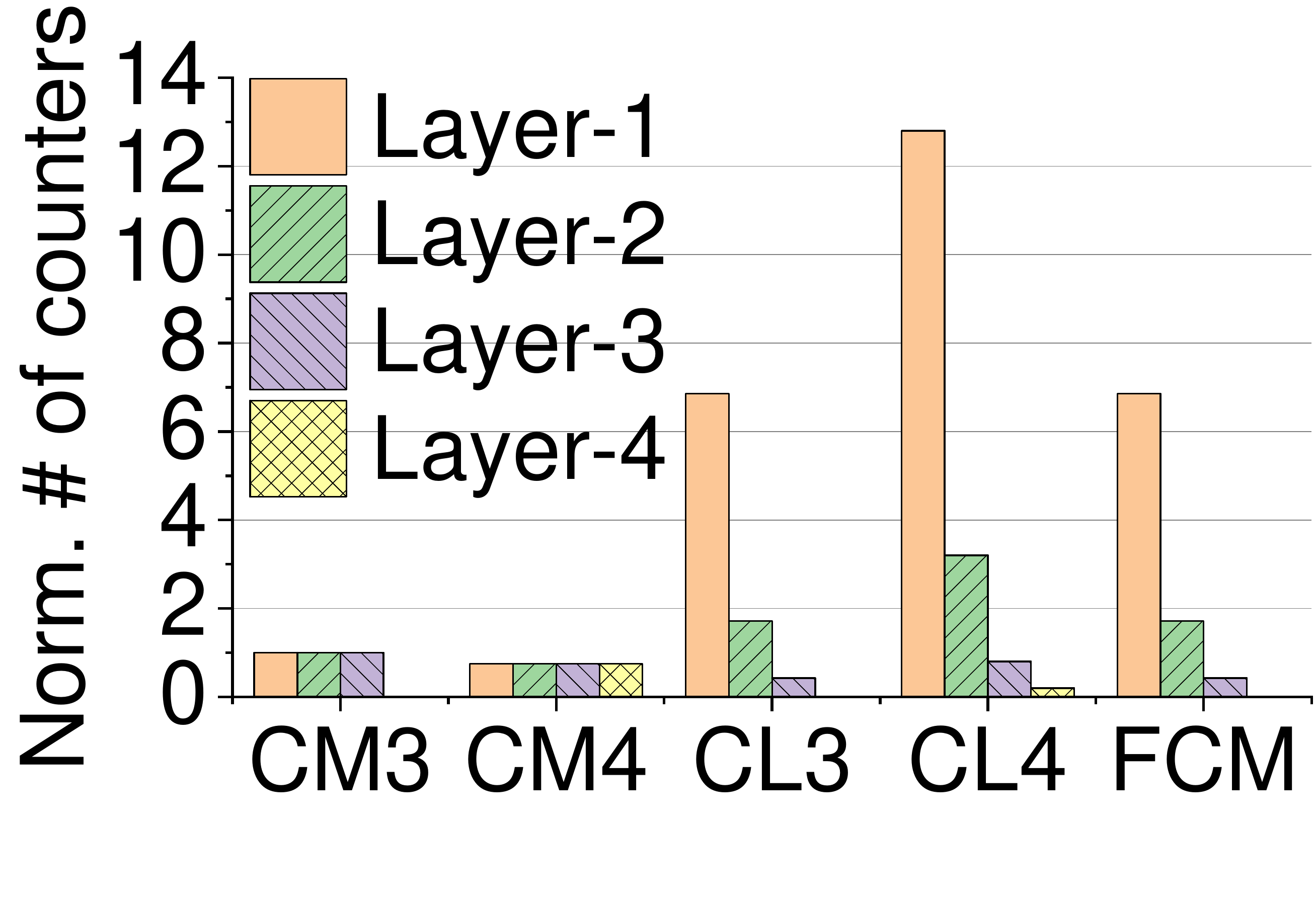}}
    \subfigure[\ours{} sketch]{\includegraphics[width=0.33\textwidth]{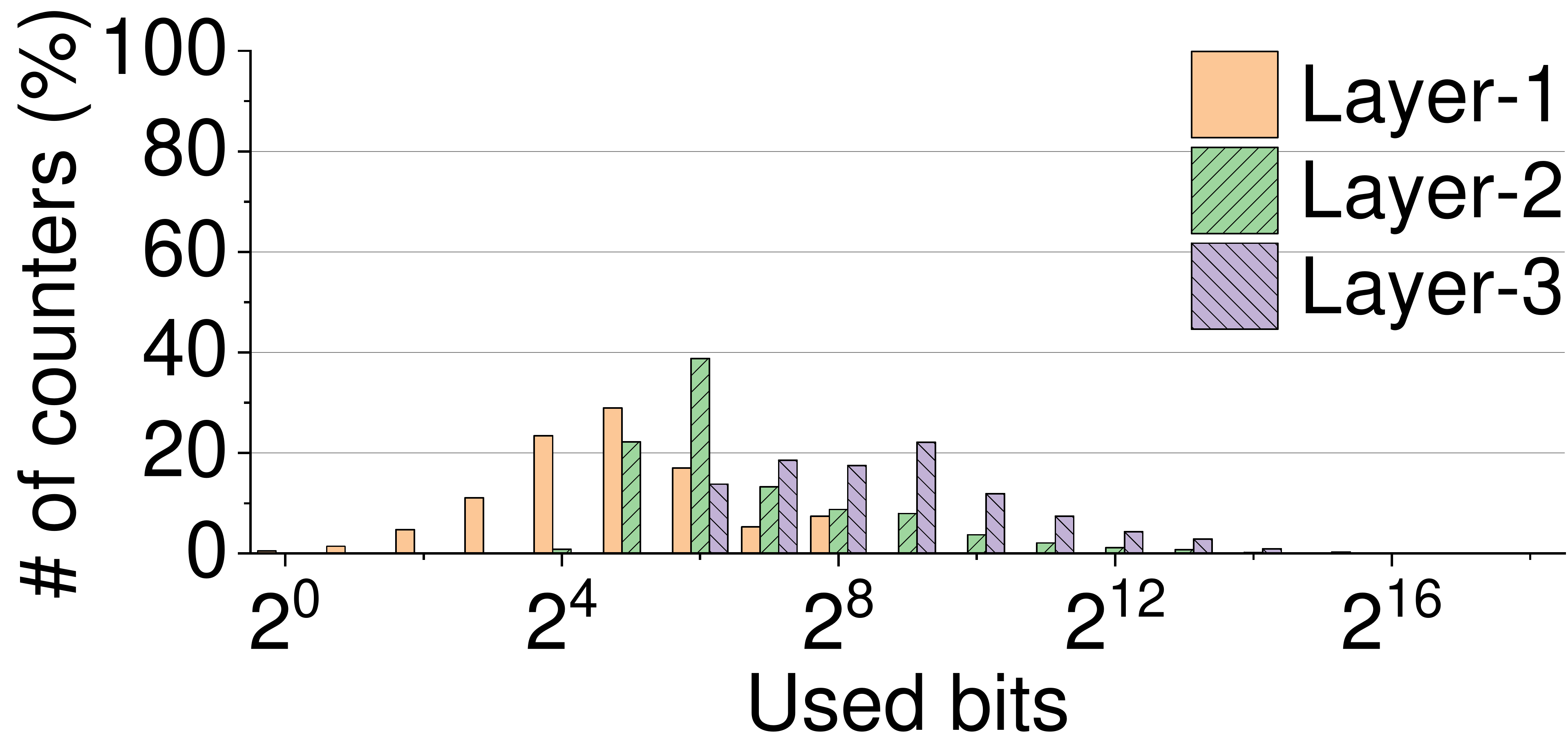}}
    \subfigure[FCM sketch]{\includegraphics[width=0.33\textwidth]{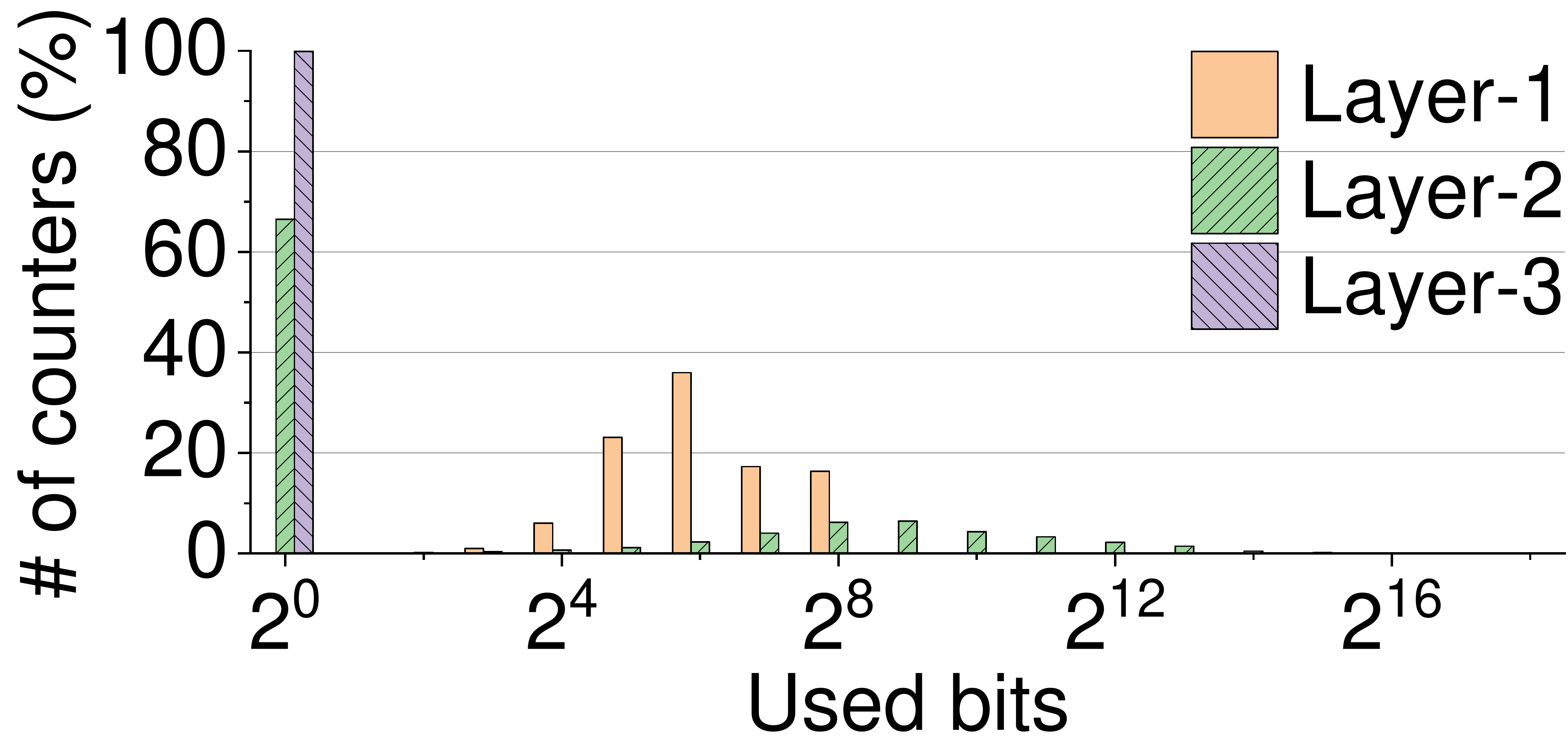}}
    \caption{Layer-wise comparison of CM, CL, and FCM in terms of the number of counters and bit-wise counter utilization. CL3 is \clmu{} with a three-layer setting ($d$=3) and CL4 is with a four-layer setting ($d$=4). Both CL3 and CL4 set the expansion factor as $r=4$. FCM uses two 4-ary trees.
		}~\label{fig:analysis_counter_distribution}
\end{figure*}

\begin{figure*}[t]
    \centering
    \includegraphics[width=0.38\textwidth]{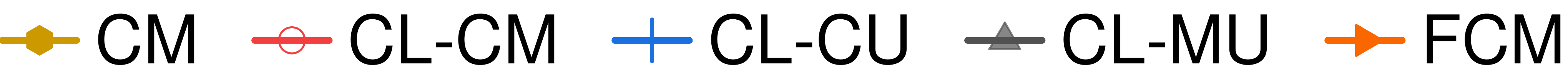}
    \\
    \subfigure[Mouse flows]{\includegraphics[width=0.25\textwidth]{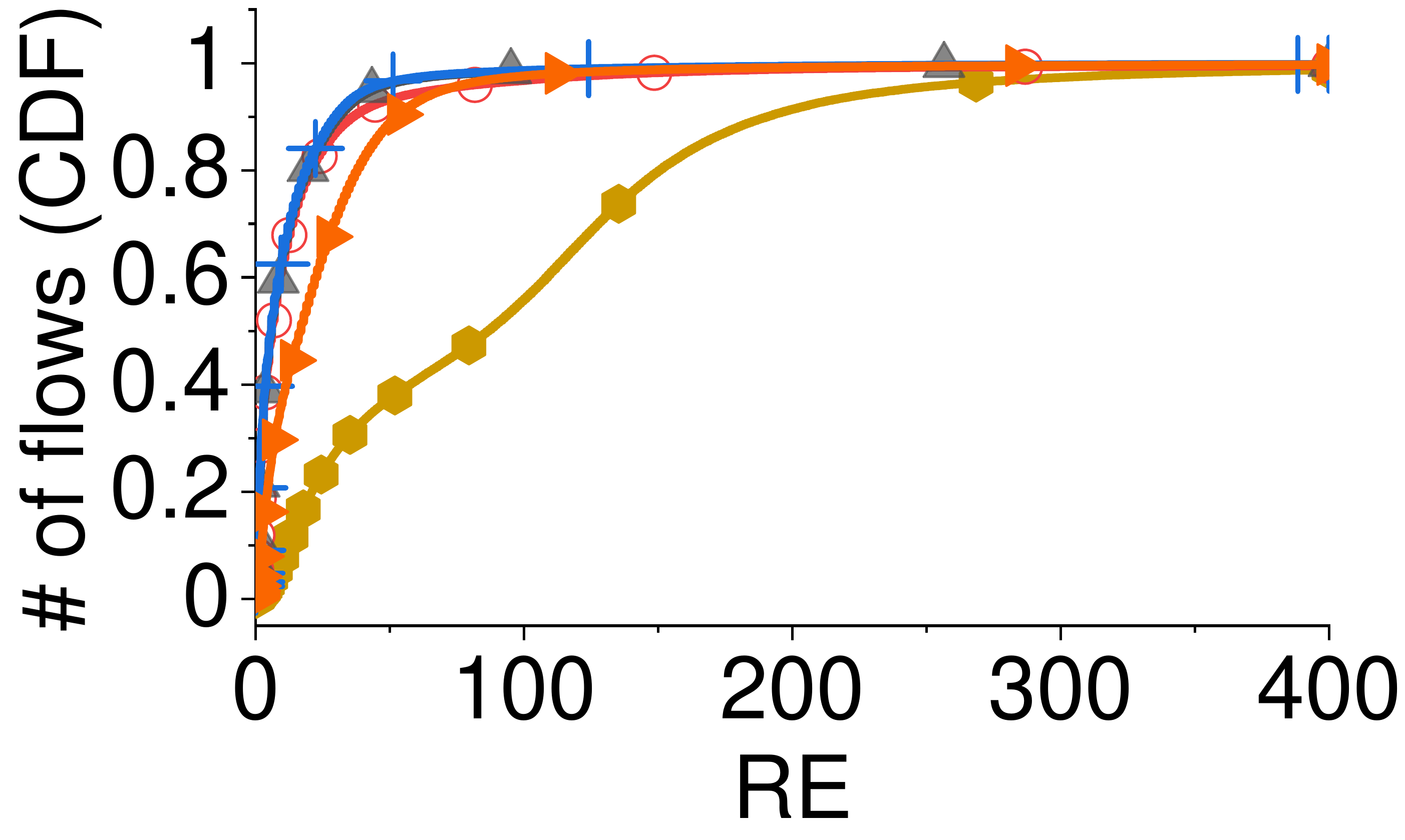}}\hspace{10mm}
    \subfigure[Medium-sized flows]{\includegraphics[width=0.25\textwidth]{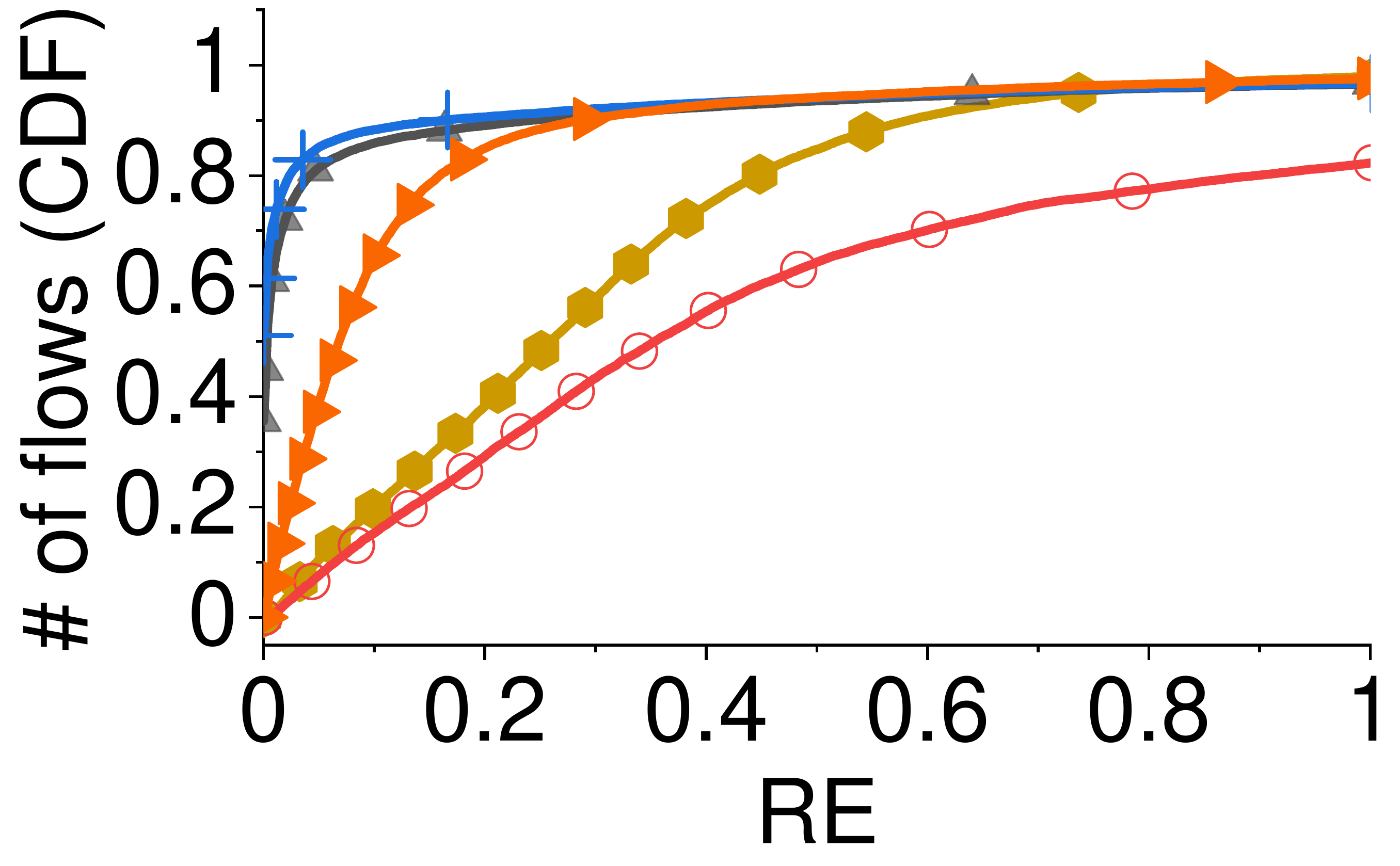}}\hspace{10mm}
    \subfigure[Elephant flows]{\includegraphics[width=0.25\textwidth]{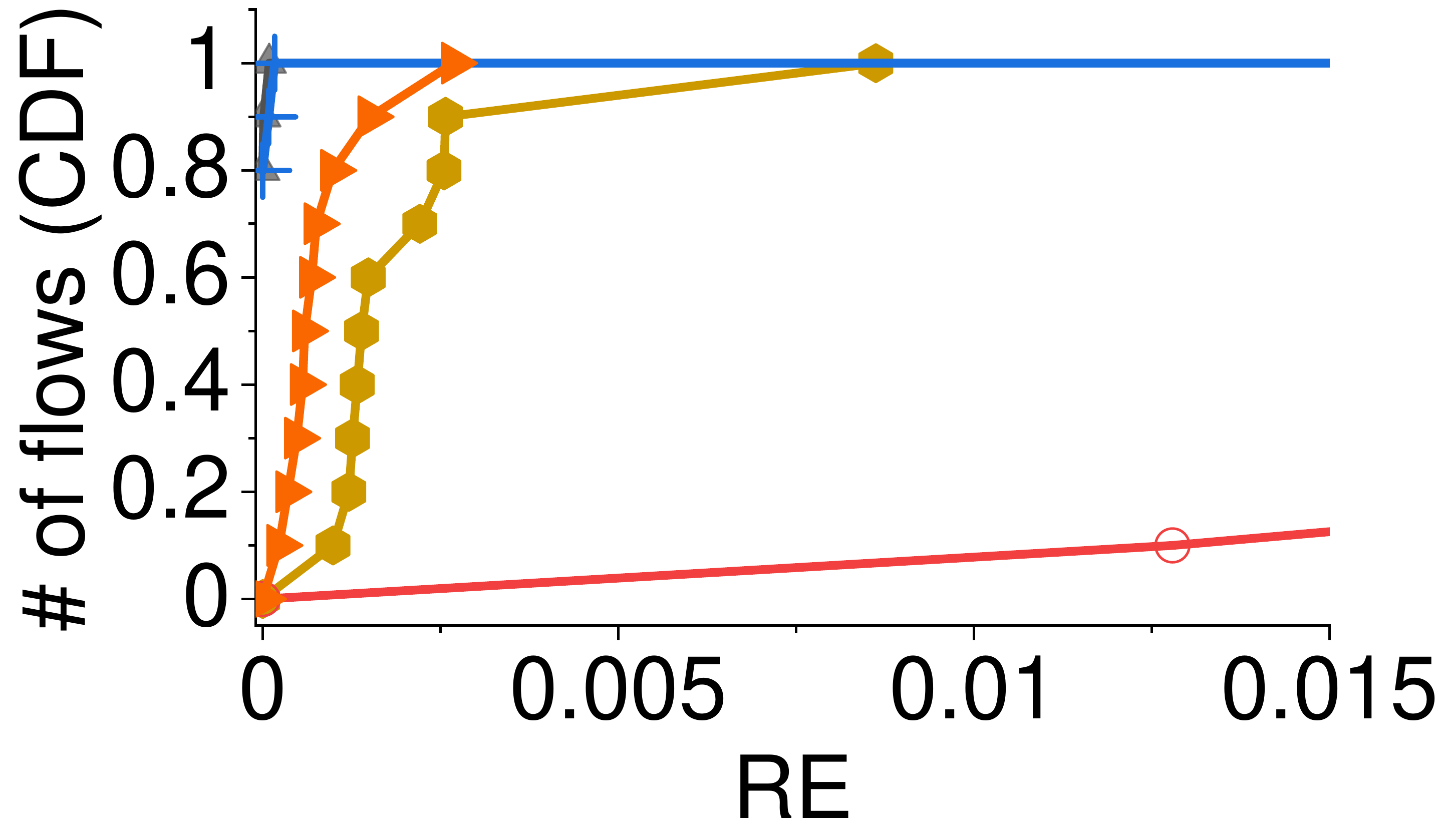}}
    \caption{CDF of relative errors for (a) mouse flows (1$\sim$254), (b) medium-sized flows (255$\sim$65534), and (c) elephant flows (65535$\sim$) by varying sketches, using one-minute CAIDA trace with 0.6 MB memory and $d=3$ in all schemes.
		}~\label{fig:minimum_update}
\end{figure*}

\subsection{Evaluating Design Choices of \ours{}}~\label{subsec:CL_eval}
To begin with, we provide detailed analysis and discuss the advantages of \ours{} from four aspects: Split Counter, Cross-layer Update, Minimum Update, and flow survival rate. Moreover, we compare \ours{} with the standard \cmin{} (CM), our baseline approach (\clcm{}), our optimal version (\clcu{}), our near-optimal and ASIC-friendly sketch (\clmu{} or CL), and the state-of-the-art work (FCM sketch).

\BfPara{(1) More counters}By splitting the counters, \ours{} can benefit from a larger number of counters to reduce the intra-layer noise by using memory more efficiently. To confirm that, we first count the number of counters in various schemes as shown in Fig.~\ref{fig:analysis_counter_distribution}(a). The figure shows the relative number of counters in CL3, CL4, CM4, and FCM against the number of counters in each layer of CM3, given the same amount of memory. The normalized number of counters of layer-1, layer-2, and layer-3 in CL3, FCM is 3.86, 1.71, and 0.43, respectively. The normalized number of counters of each layer in CL4 is 12.8, 3.2, 0.8, and 0.2, from bottom to top.
Due to this structure, the number of counters available for mouse flows, which accounts for a larger proportion of the Zipfian distribution, increases and the accuracy is improved.

\BfPara{(2) Cross-layer Update} As  can be seen in Fig.~\ref{fig:analysis_counter_distribution}(a), FCM sketch also has the same number of counters as \ours{}. However, FCM rarely uses layer-2 and layer-3 for the small flows due to the cascade counting method, whereas \ours{} sketch uses all layers with the cross-layer update strategy. Fig.~\ref{fig:analysis_counter_distribution}(b),(c) shows the number of bits in use per counter after counting the one-minute CAIDA dataset in 0.6 MB sketches. For \ours{} sketch, 83.9\% of the counters in layer-2 and 49.9\% of counters in layer-3 have values of less than 8 bits. This means that mouse flows (\ie $<2^8-1$) can be decoded in layer-2 and layer-3, allowing mouse flows to survive at the upper layers. As a result, \ours{}'s mouse flow estimation is more accurate than that of CM and FCM, as shown in Fig.~\ref{fig:minimum_update}(a). In contrast, in FCM sketch, 66.5\% of counters in layer-2 and 99.9\% of counters in layer-3 were not used at all. At the same time, \ours{} has ARE of 15.72 whereas FCM sketch has ARE of 26.45, showing the cross-layer update strategy that makes \ours{} more accurate.

\BfPara{(3) Minimum Update} To show the effect of {\em Minimum Update}, we compared the noise level (\ie RE) of CM3, CL3-CM, CL3-MU, CL3-CU, and FCM for one-minute CAIDA dataset with 0.6 MB of memory.
Fig.~\ref{fig:minimum_update}(a) shows that CL3-CM, CL3-MU, CL3-CU, and FCM that with split counters have less noise in mouse flows (1$\sim$254) than CM3 with the flat counters, which shows the advantage of the split counter approach.
However, for the medium-sized flows (255$\sim$65,534), CL3-MU and CL3-CU have substantially lower relative error than other sketches adopting the same split counter, including FCM as shown in Fig.~\ref{fig:minimum_update}(b). Similarly, for elephant flows (65,535$\sim$), CL3-MU and CL3-CU outperform the other split counter-based sketches including FCM, significantly, as shown in Fig.~\ref{fig:minimum_update}(c). This experiment confirms that the cross-layer update-based Minimum/Conservative Update is the key factor contributing the superiority in estimation of the medium-sized/elephant flows. 

\begin{figure*}[t]
    \centering
    \subfigure[Mouse flow(size $<$ 255)]{\includegraphics[width=0.37\textwidth]{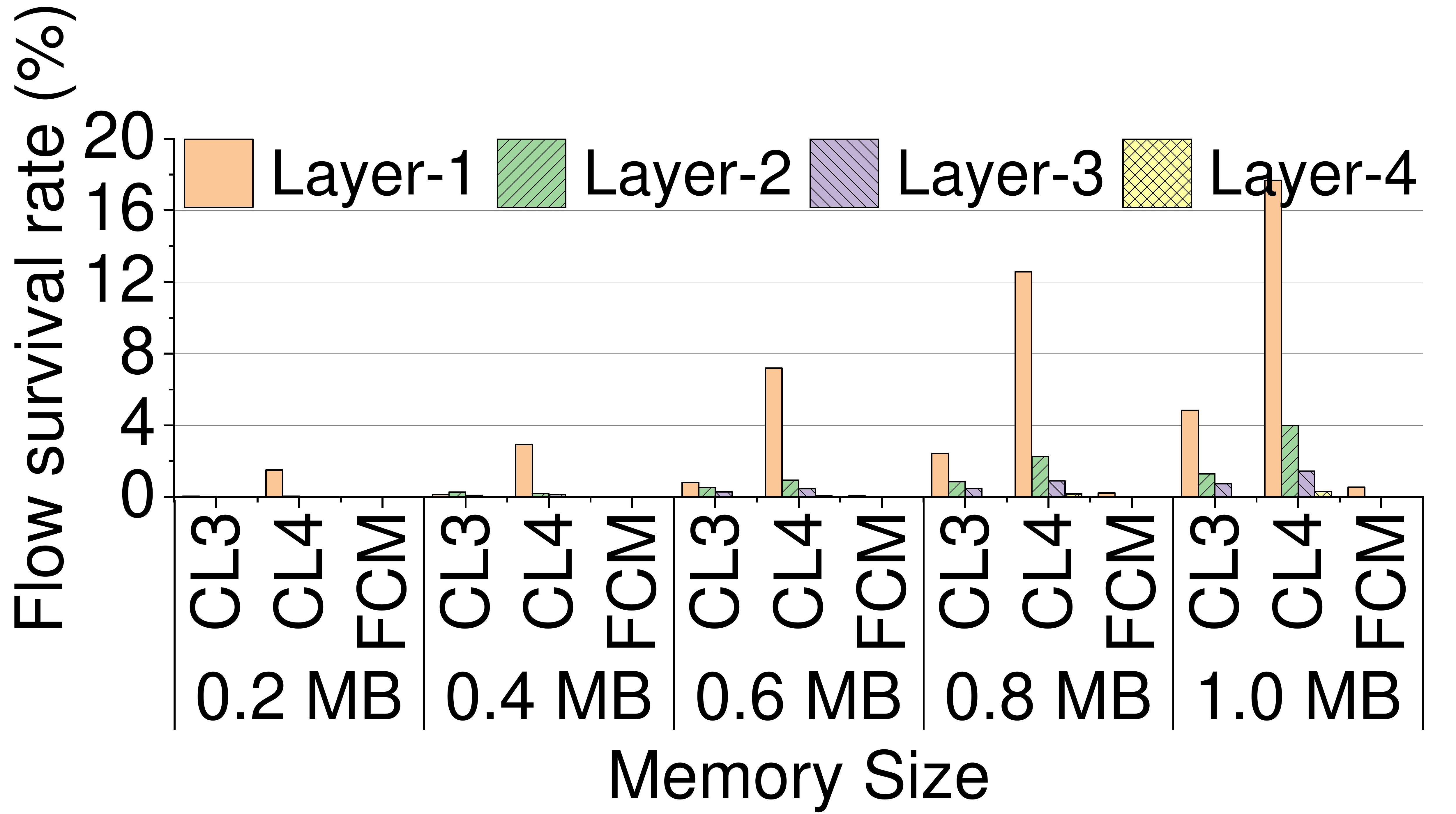}}
    \hspace{6mm}
    \subfigure[Medium-sized and elephant flows (size$\geq$255)]{\includegraphics[width=0.37\textwidth]{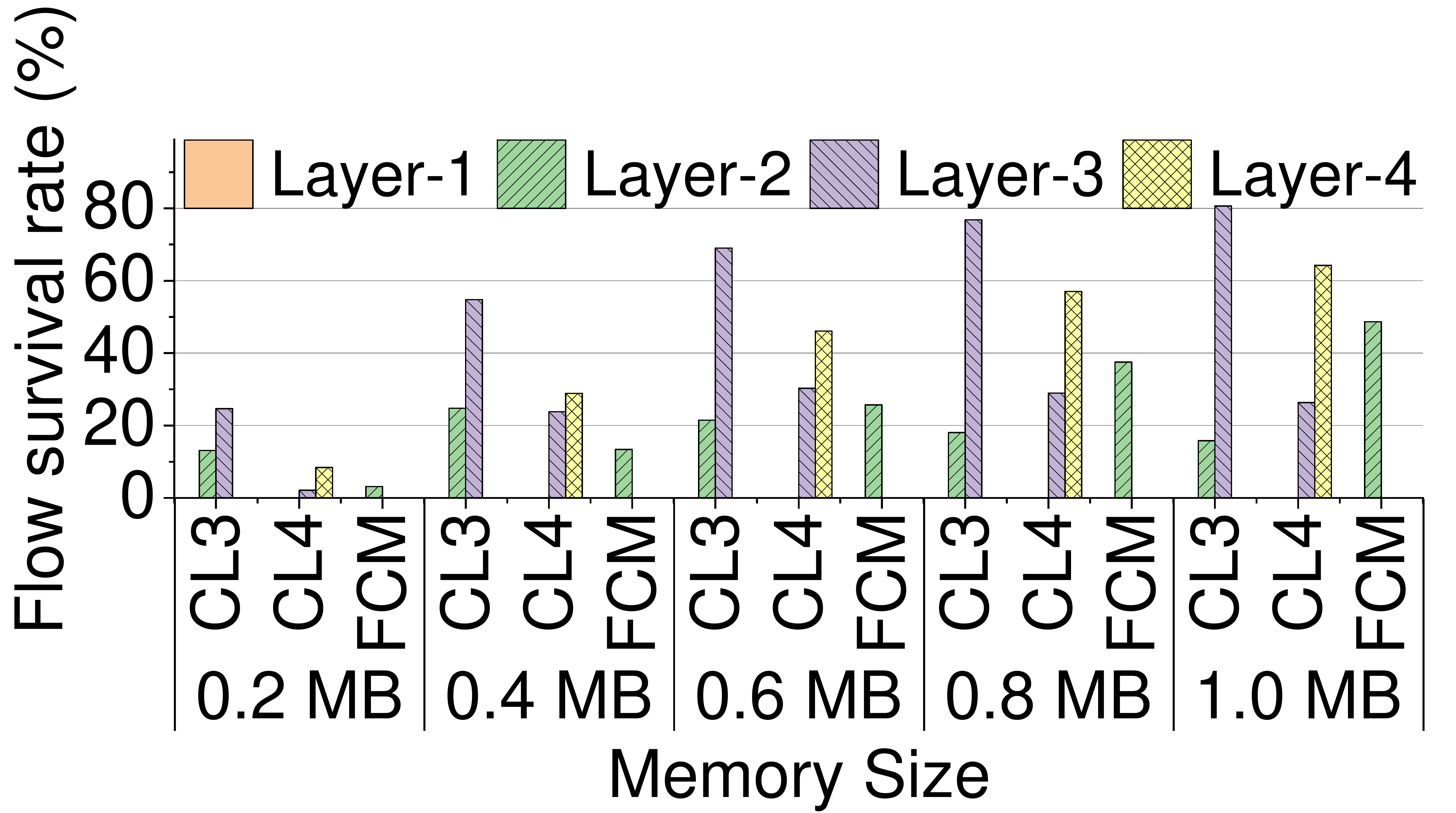}}
    \caption{Flow survival rate comparison: a mouse flow is considered as survived if the estimated relative error is smaller than 0.1, a medium-sized flow (255$\sim$65,534) is below 0.05, and elephant flow (65,535$\sim$) is below 0.01. 
		}~\label{fig:survival_rate}
\end{figure*}

\BfPara{(4) More Survived Flows (Synergy Effect)}\label{sec:synergy} One of the strong indicators of estimation accuracy is the flow survival rate (FSR) defined by a fraction of the flows whose RE is lower than a specific criterion ($RE<0.1$ for mouse flows, $RE<0.05$ for mid-sized flows, and $RE<0.01$ for elephant flows). To find out how many flows survive in each sketch, we measured the flow survival rate of CL3-MU, CL4-MU, and FCM sketch for mouse flows and elephant flows separately. 
\ours{} may decode mouse flows in all layers owing to the cross-layer update strategy, while FCM sketch can decode only at layer-1 owing to its overflow-based update strategy, which is shown by the survival rate at each layer in Fig.~\ref{fig:survival_rate}(a). As shown in the figure, the 
overall survival rates of the mouse flows in CL3-MU, CL4-MU, and FCM range from 0.08\% to 6.87\%, from 1.57\% to 23.43\%, and from 0 to 0.56\%, respectively. 
In Fig.~\ref{fig:survival_rate}(b), we observe that 37.78\% to 96.42\% of the medium-sized and elephant flows survive in CL3-MU, 10.51\% to 90.61\% survive in CL4-MU, and 3.20\% to 48.72\% survive in FCM.
This result indicates that much more flows are likely to survive in \ours{}
that combines both Split Counter and Minimum Update, 
making \ours{}'s estimation more accurate than FCM over all ranges even with a smaller memory footprint. For example, CL4-MU with 0.2 MB memory (1.57\% FSR) works better than FCM with 1 MB memory (0.56\% FSR) for mouse flows, and CL3-MU with 0.4 MB memory (79.53\% FSR) is better than FCM with 1 MB memory (48.72\% FSR) for medium-sized and elephant flows.

\subsection{Configuring Parameters}\label{sec:param_config}
\begin{figure*}[t]
    \centering
    \subfigure[ARE varying $r$]{\includegraphics[width=0.25\textwidth]{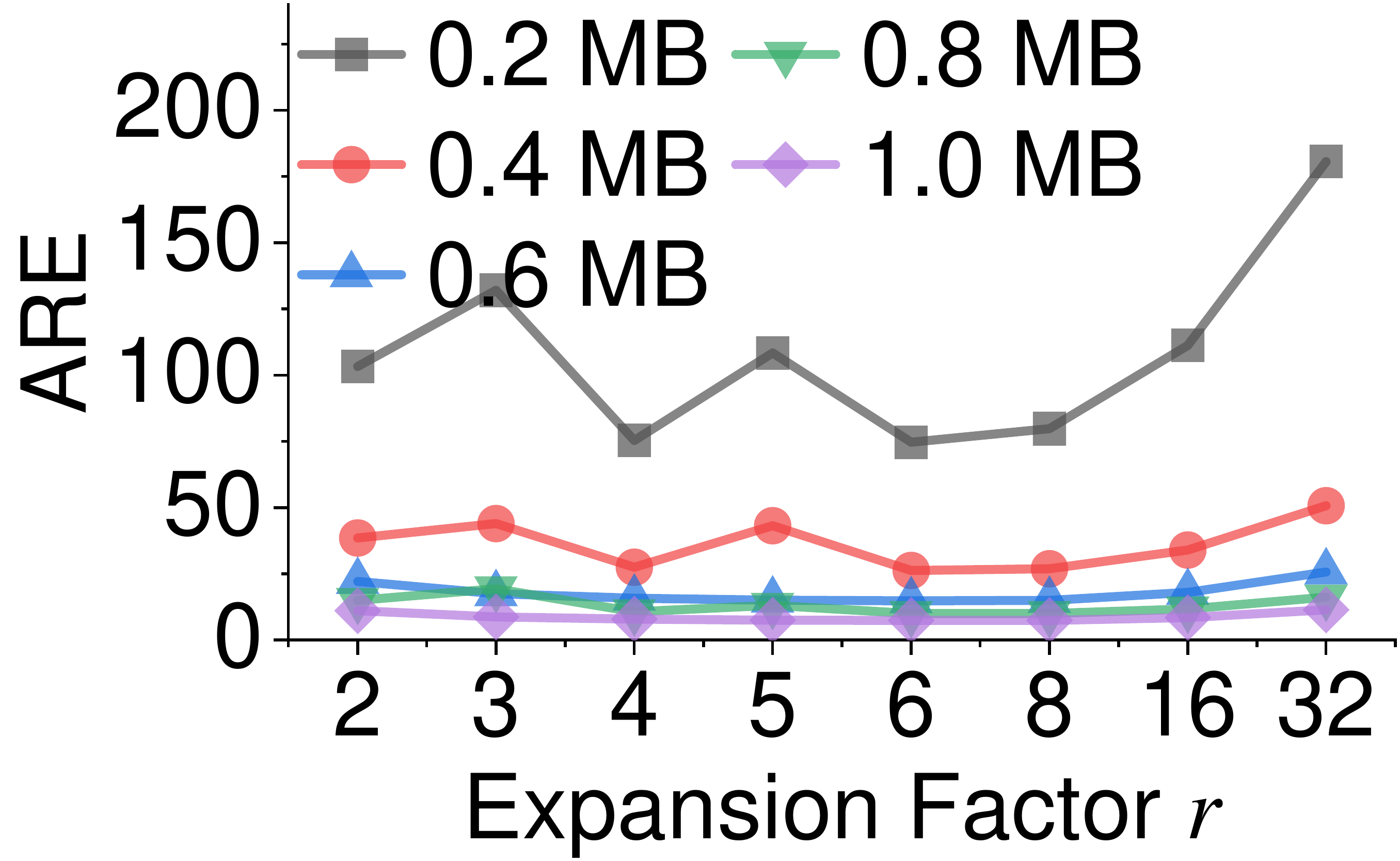}}\hspace{10mm}
    \subfigure[Norm. \# of counters varying $r$]{\includegraphics[width=0.25\textwidth]{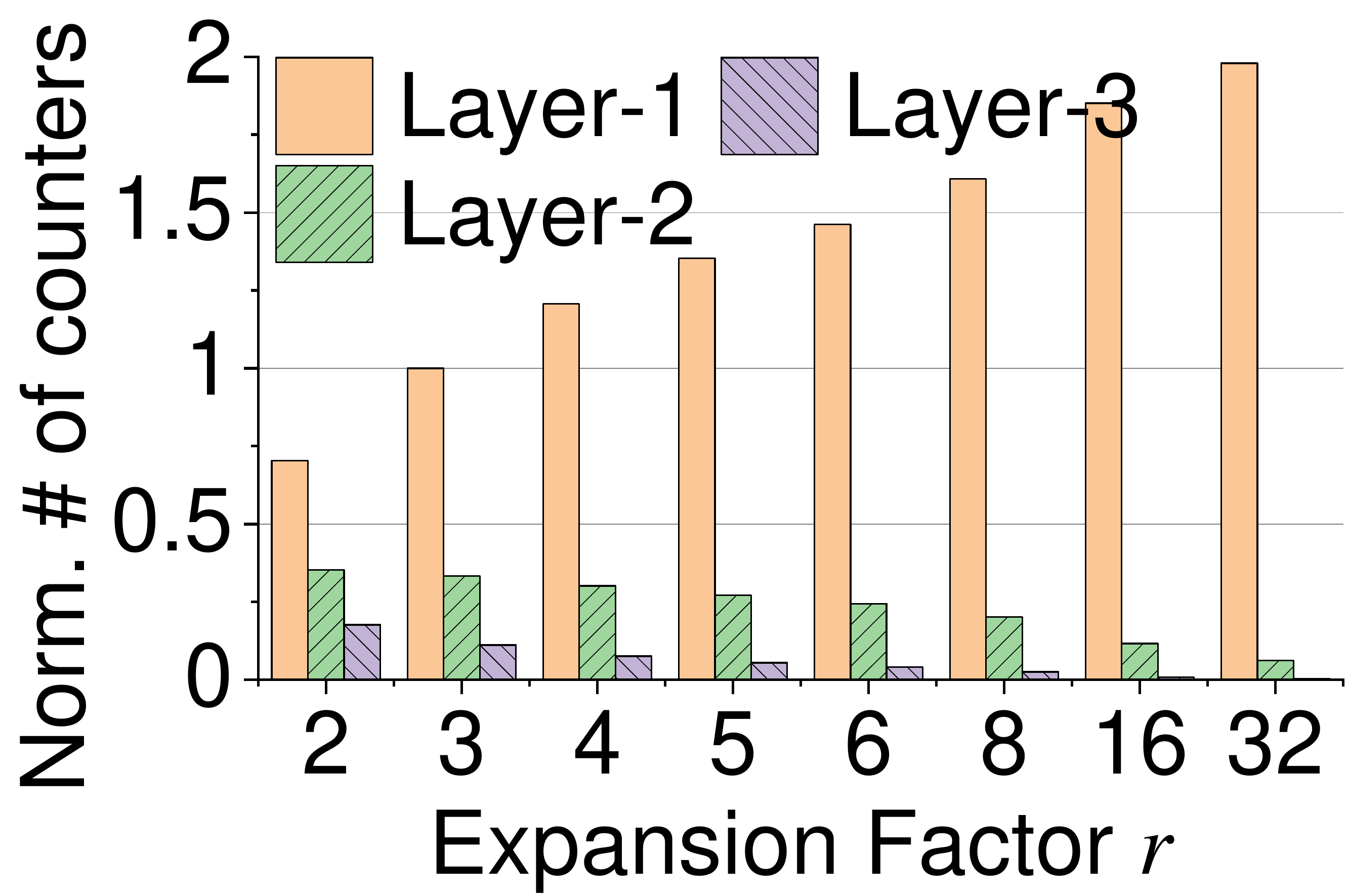}}\hspace{10mm}
    \subfigure[ARE varying $d$]{\includegraphics[width=0.25\textwidth]{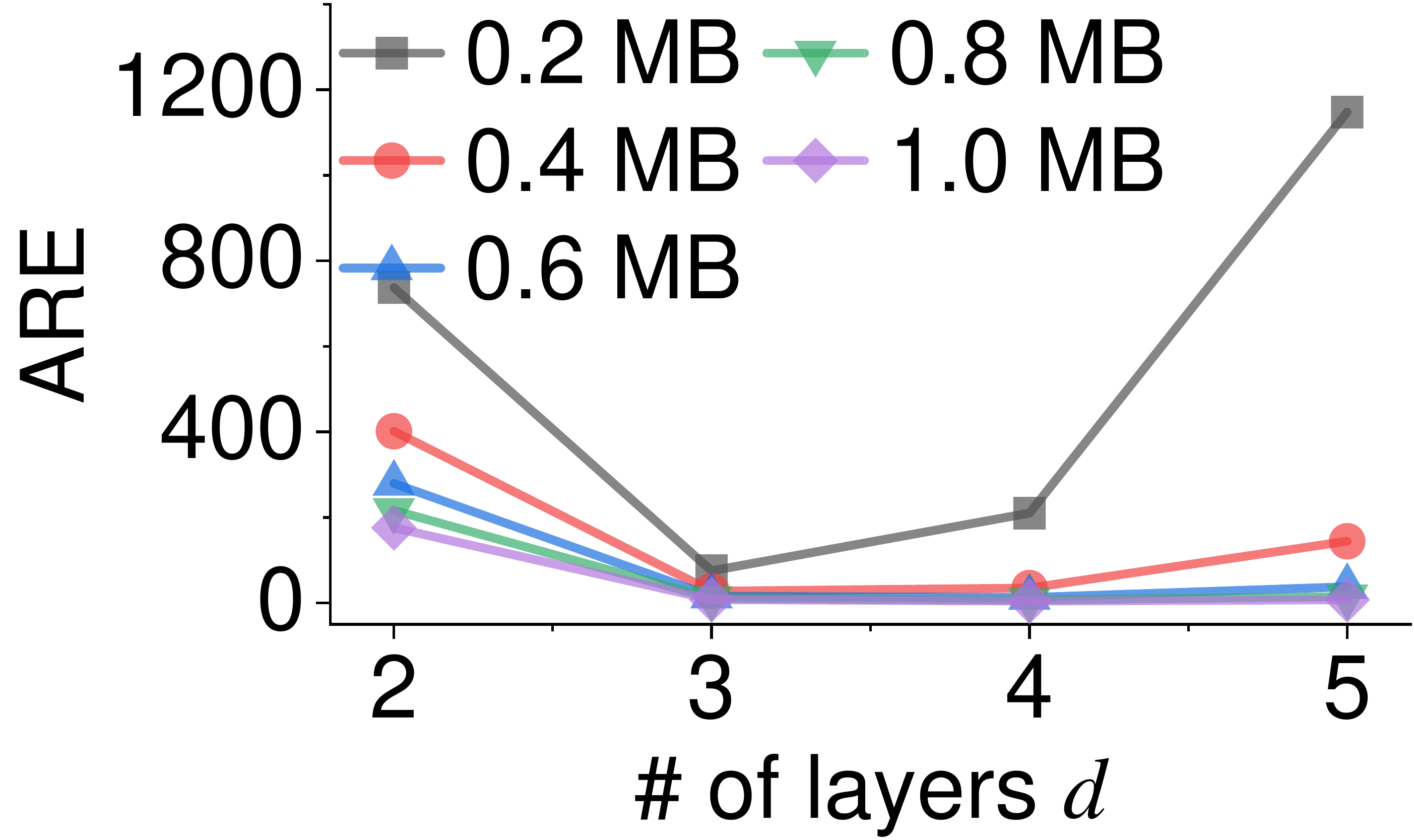}}
    \caption{
		     (a) and (c) show accuracy of Count-Less with different configurations is shown, varying the number of layers $d$ and the expansion factor $r$. (b) presents the number of counters normalized by layer-1 in CL3 with $r=3$, varying $r$ with the same size of the memory.
		}~\label{fig:param}
\end{figure*}
In this section, the accuracy of \ours{} with different configurations is examined by varying the number of layers $d$ and the expansion factor $r$. To evaluate the accuracy, we used a one-minute CAIDA trace with a memory size of 0.2 MB to 1 MB.
Fig.~\ref{fig:param}(a) shows the ARE of CL3-MU with various values of the expansion factor $r$. As shown in Fig.~\ref{fig:param}(a), CL3-MU with $r=6$ or $r=8$ has the lowest ARE regardless of the memory size, and the accuracy of $r=4$ is comparable to it.
 Fig.~\ref{fig:param}(b) shows the number of counters normalized by layer-1 in CL3 with $r=3$ while varying $r$ with the same size of the memory. \ours{} is shown to have more accurate results as $r$ increases, because the total number of counters increases proportionally to the number of counters in layer-1, as shown in  Fig.~\ref{fig:param}(b). On the other hand, the number of counters in layer-2 and layer-3 becomes smaller, with the normalized number of counters at $r=32$ being 0.062 and 0.0019, respectively. Because of this, larger $r$ makes CL more sensitive to the skewness of the distribution, thus becomes less accurate.
Fig.~\ref{fig:param}(c) shows ARE of \clmu{} with $r=4$, varying the number of layers $d$. In the memory size of 0.2 MB and 0.4 MB, CL3-MU is the most accurate with ARE of 24.43 and 75.31. CL4-MU shows the highest accuracy with ARE of 12.96, 6.87, and 4.39, using 0.6 MB, 0.8 MB, and 1 MB of memory.
Based on the parameterization experiment, we recommend using \ours{} with $r=4$, $d=3, 4$.

\begin{figure*}[t]
    \centering
    
    \includegraphics[width=0.40\textwidth]{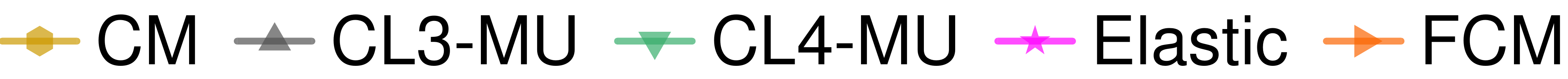}
    \\
    \subfigure[Flow Size Estimation]{\includegraphics[width=0.25\textwidth]{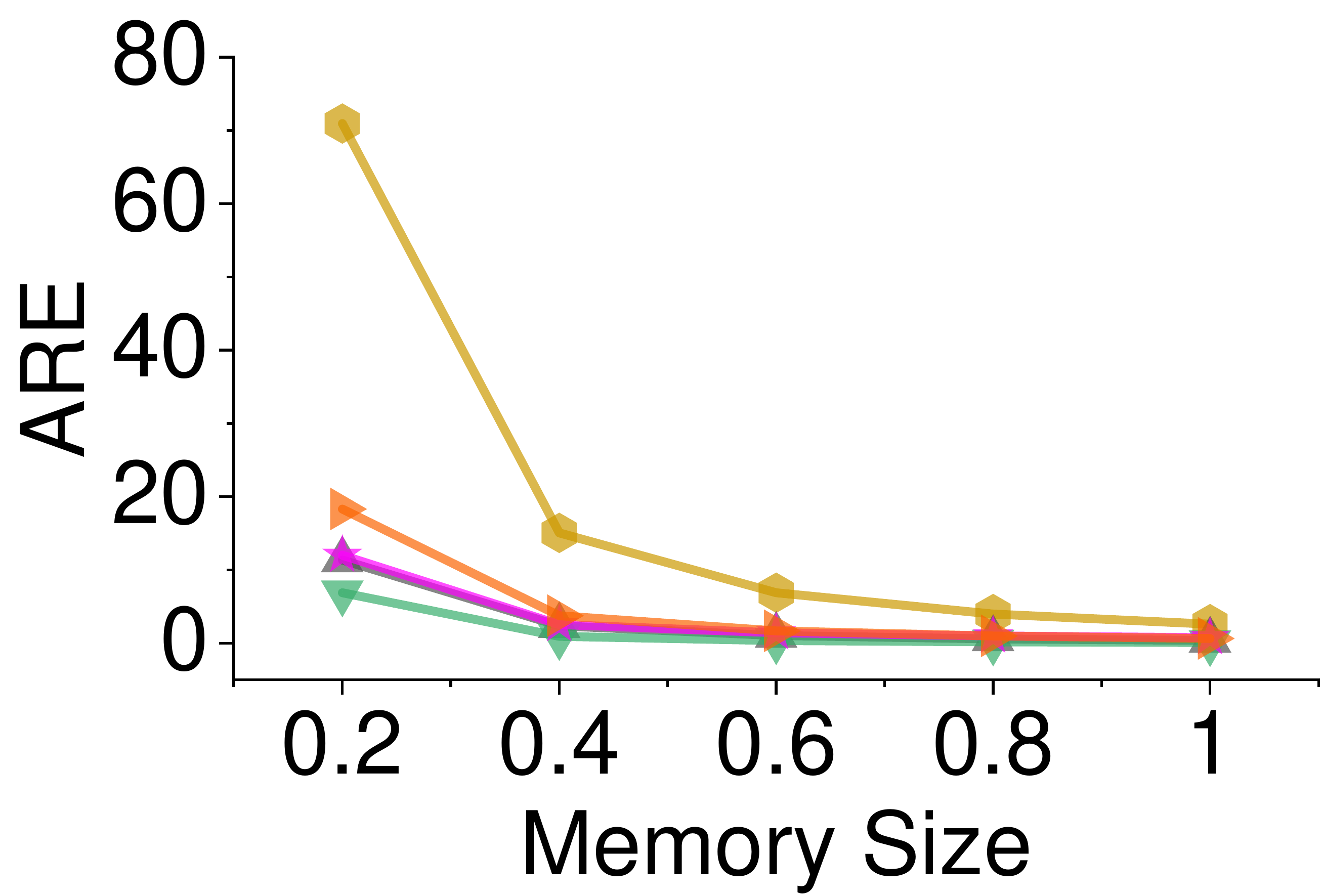}}\hspace{6mm}
    \subfigure[Cardinality]{\includegraphics[width=0.25\textwidth]{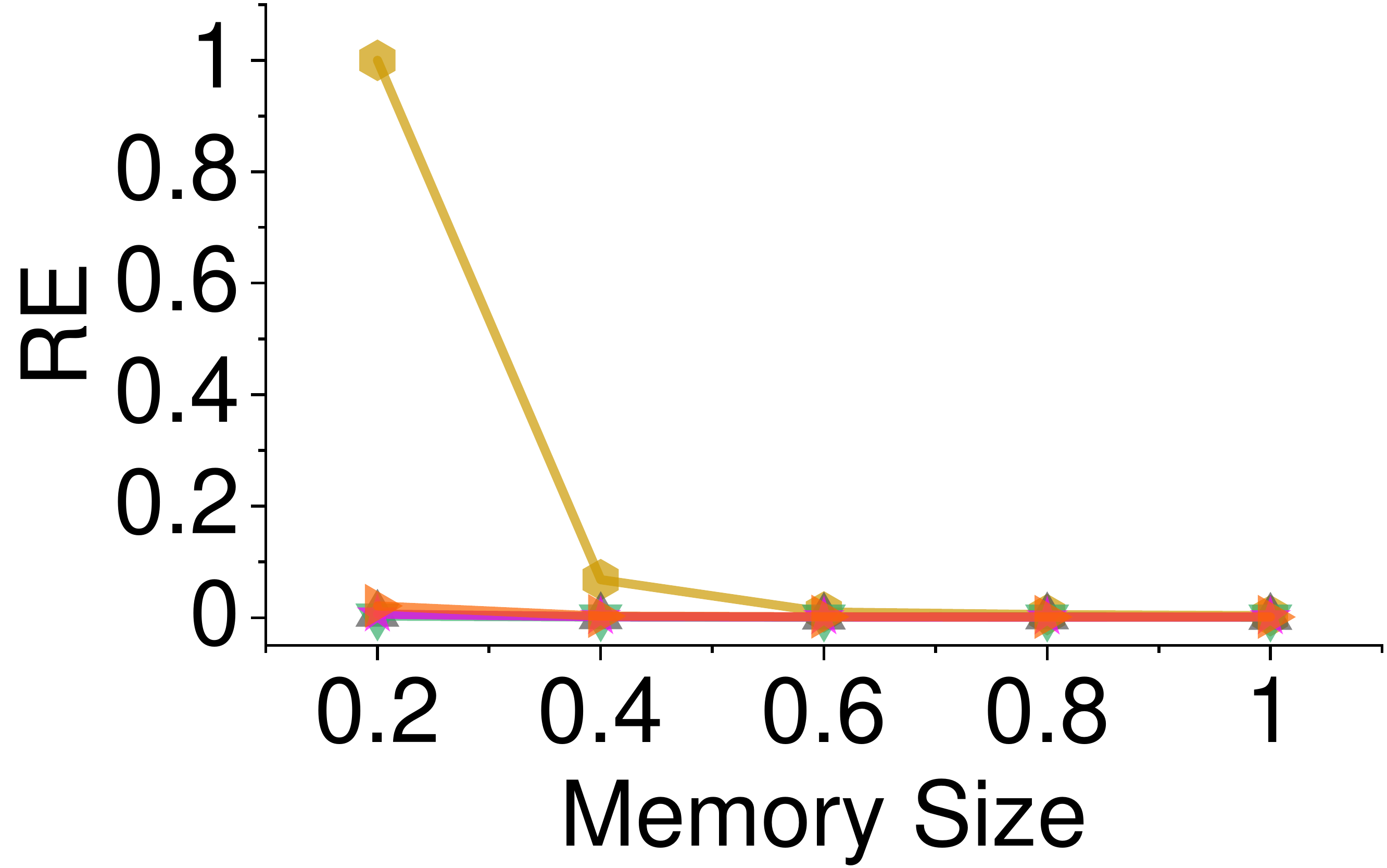}}\hspace{6mm}
    \subfigure[Flow Size Distribution]{\includegraphics[width=0.25\textwidth]{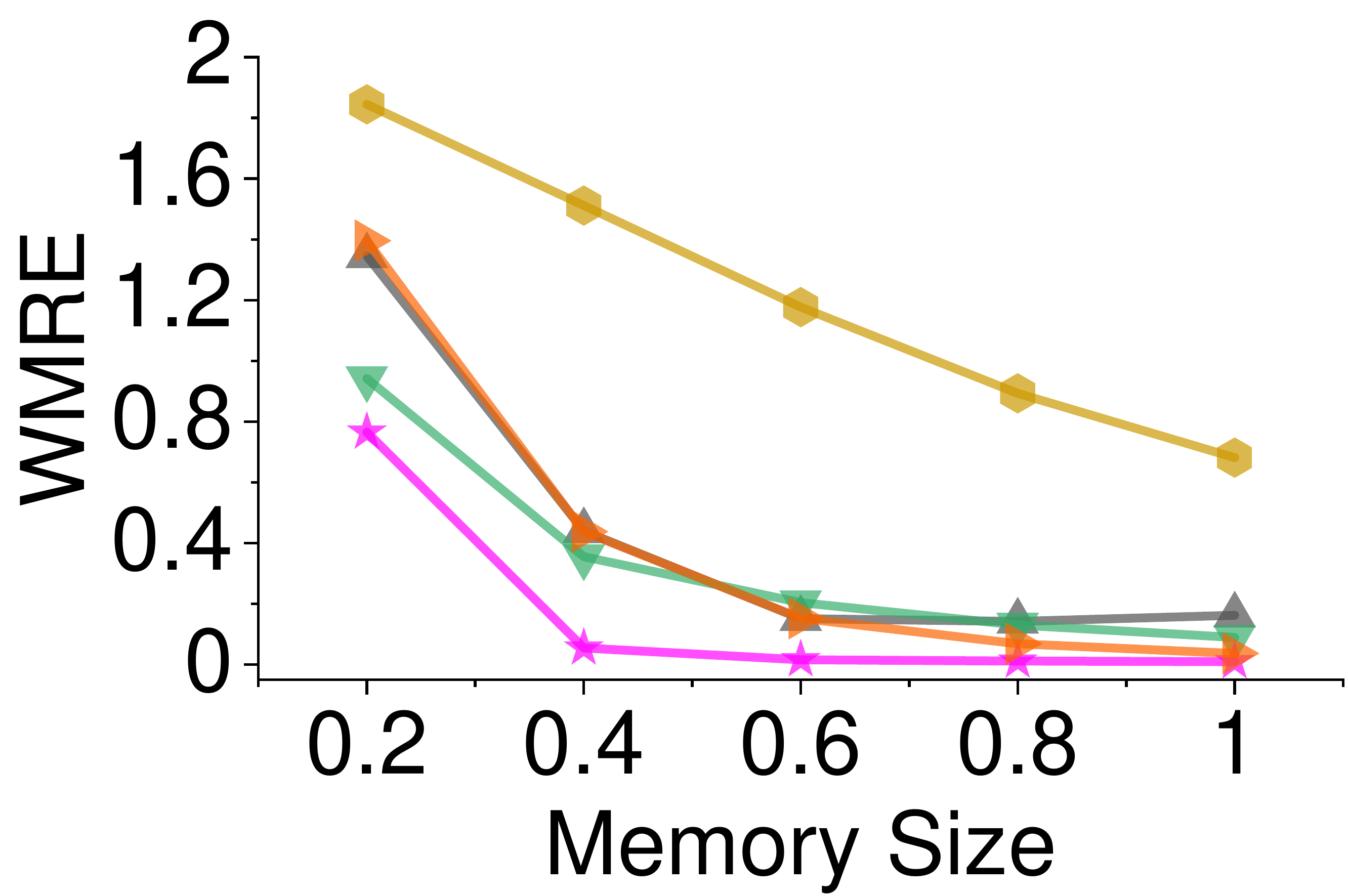}}
    \\
    \subfigure[Entropy]{\includegraphics[width=0.25\textwidth]{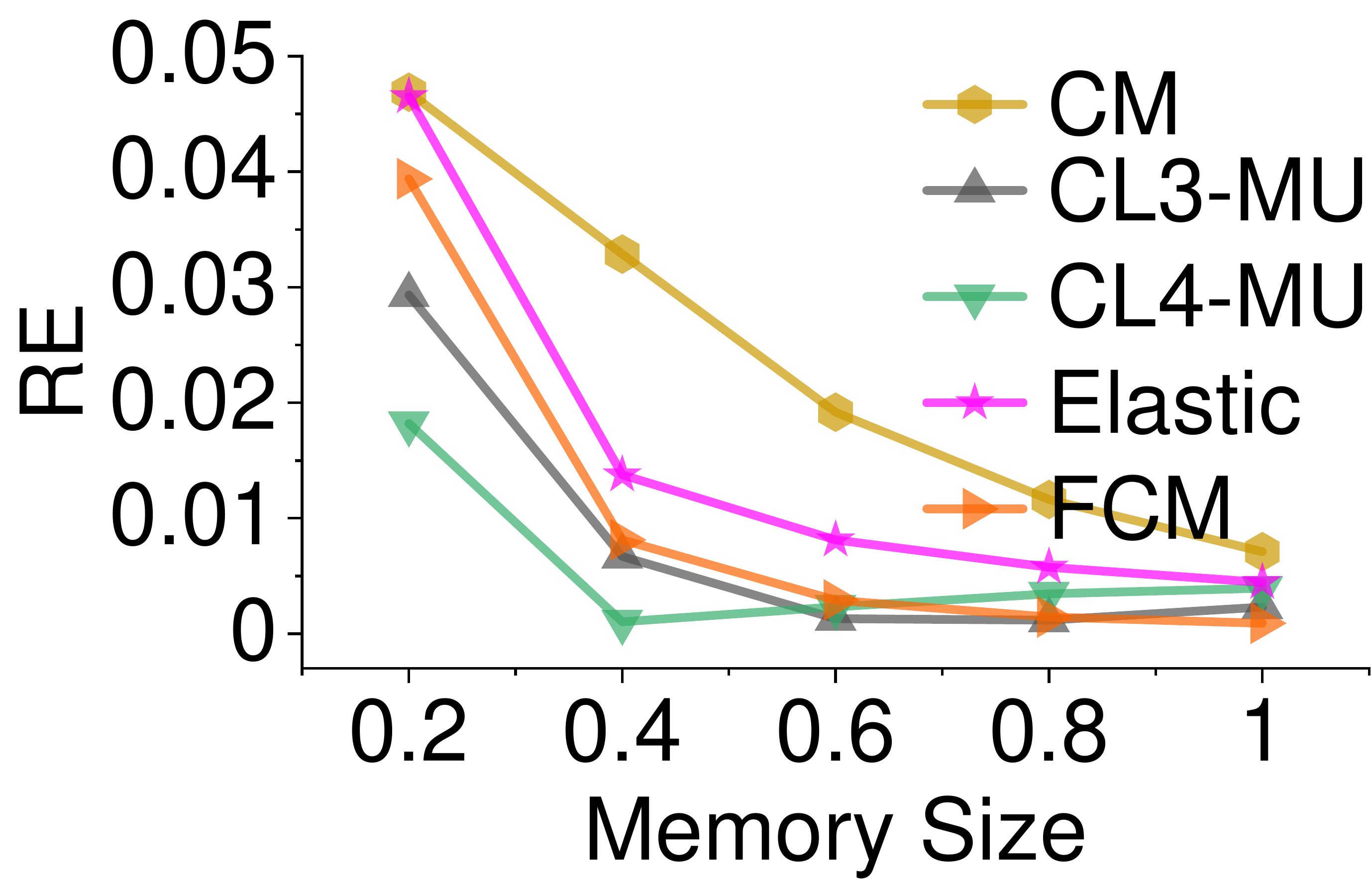}}\hspace{6mm}
    \subfigure[Heavy Hitter]{\includegraphics[width=0.25\textwidth]{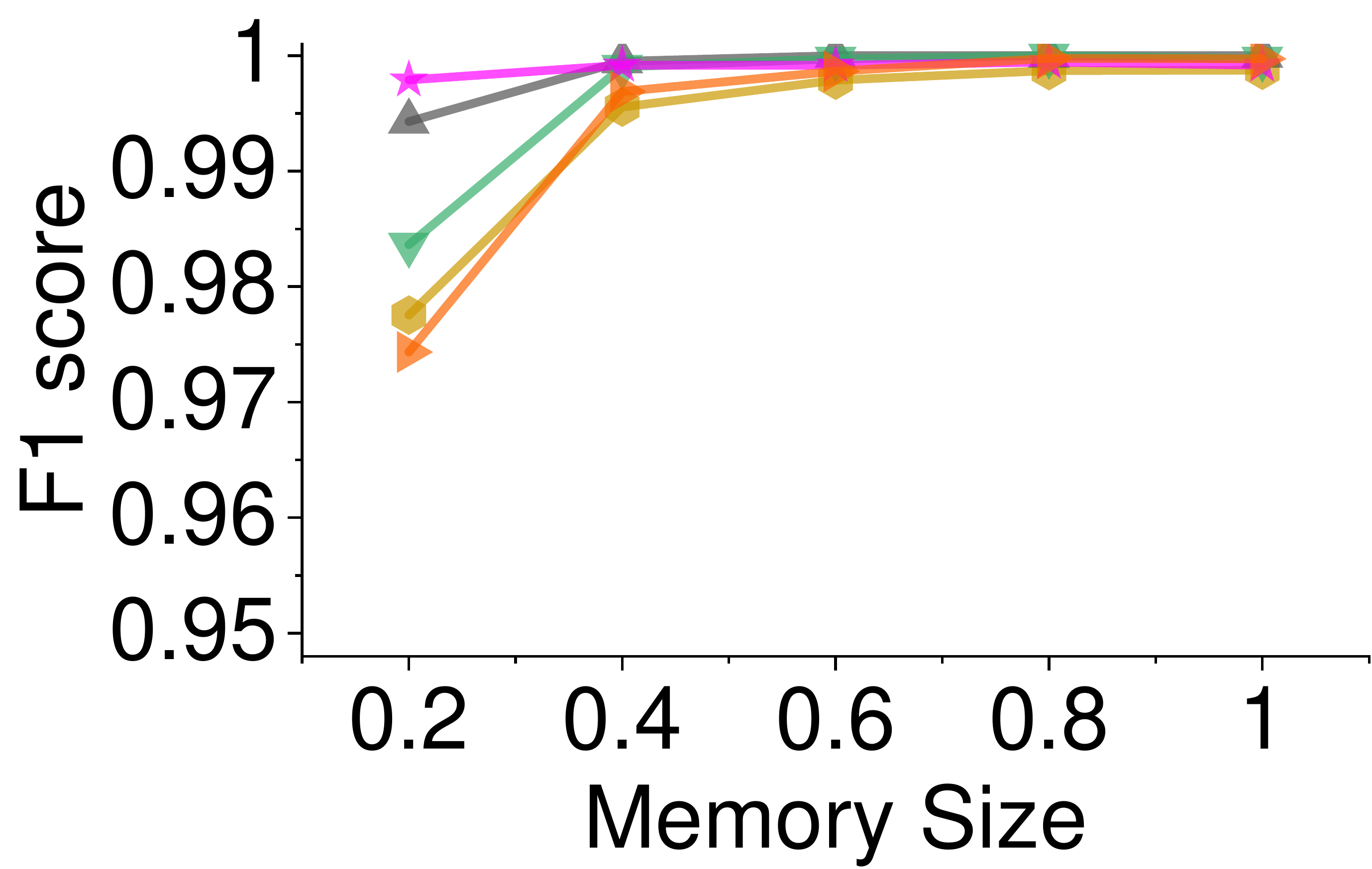}}\hspace{6mm}
    \subfigure[Heavy Changer]{\includegraphics[width=0.25\textwidth]{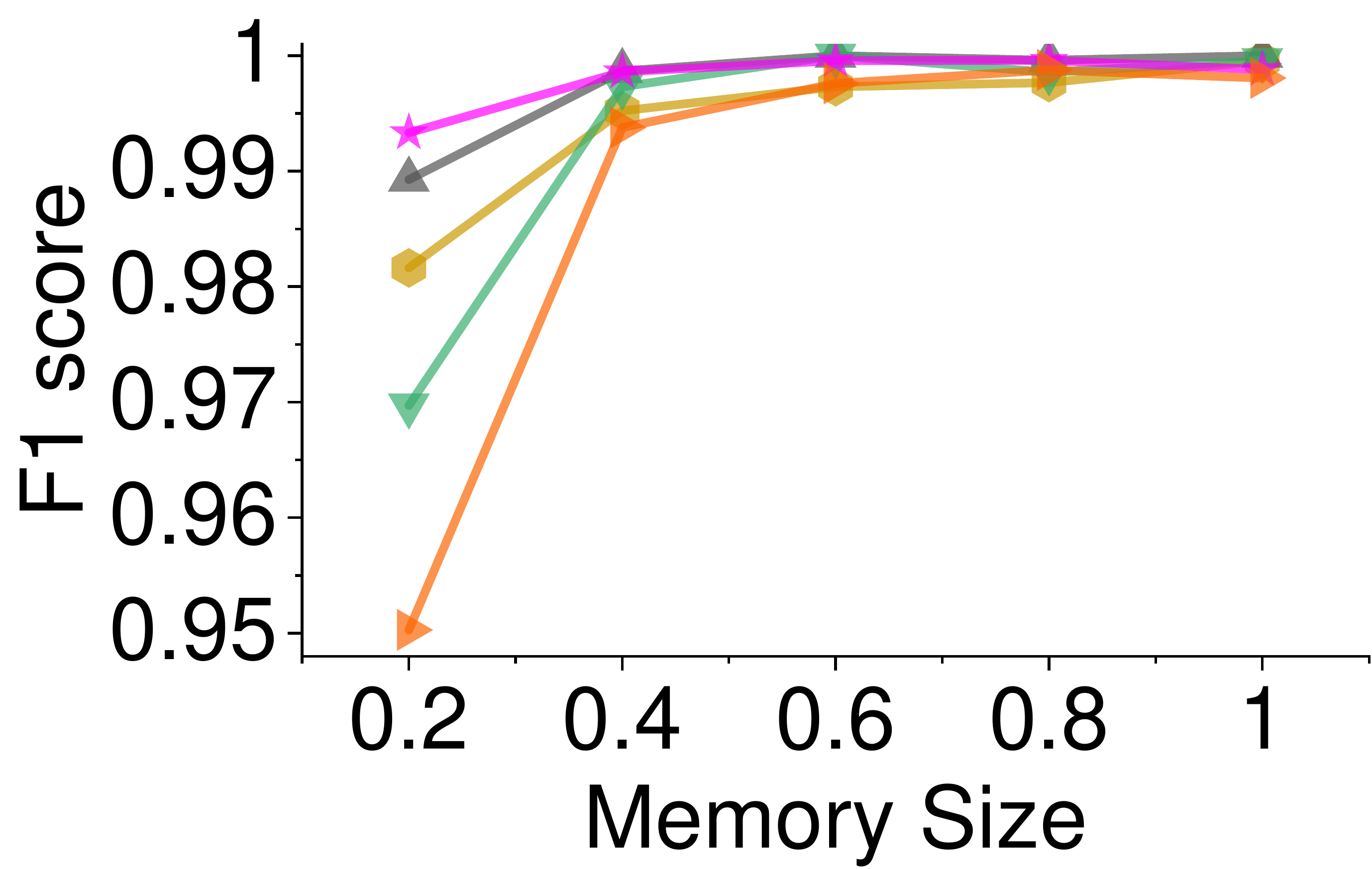}}
    \caption{
		     Accuracy comparison of ASIC-friendly \ours{} (CL-MU), \cmin{} (CM), Elastic, and FCM sketches in network traffic measurement tasks. All sketches are assigned the same memory size from 0.2 MB to 1 MB. \ours{}'s performance is given for both three- and four-layer settings (\ie CL3-MU and CL4-MU). We note that for flow size distribution and entropy, \ours{} used our lightweight algorithm to focus on in-router processing (see in Fig.~\ref{fig:dist_arr_p4}), whereas Elastic and FCM used computation-intensive MRAC for better accuracy.
		}~\label{fig:cpu_exp_1}

\end{figure*}

\subsection{Applications}~\label{subsec:application}
\begin{figure}
    \centering
    \includegraphics[width=0.43\textwidth]{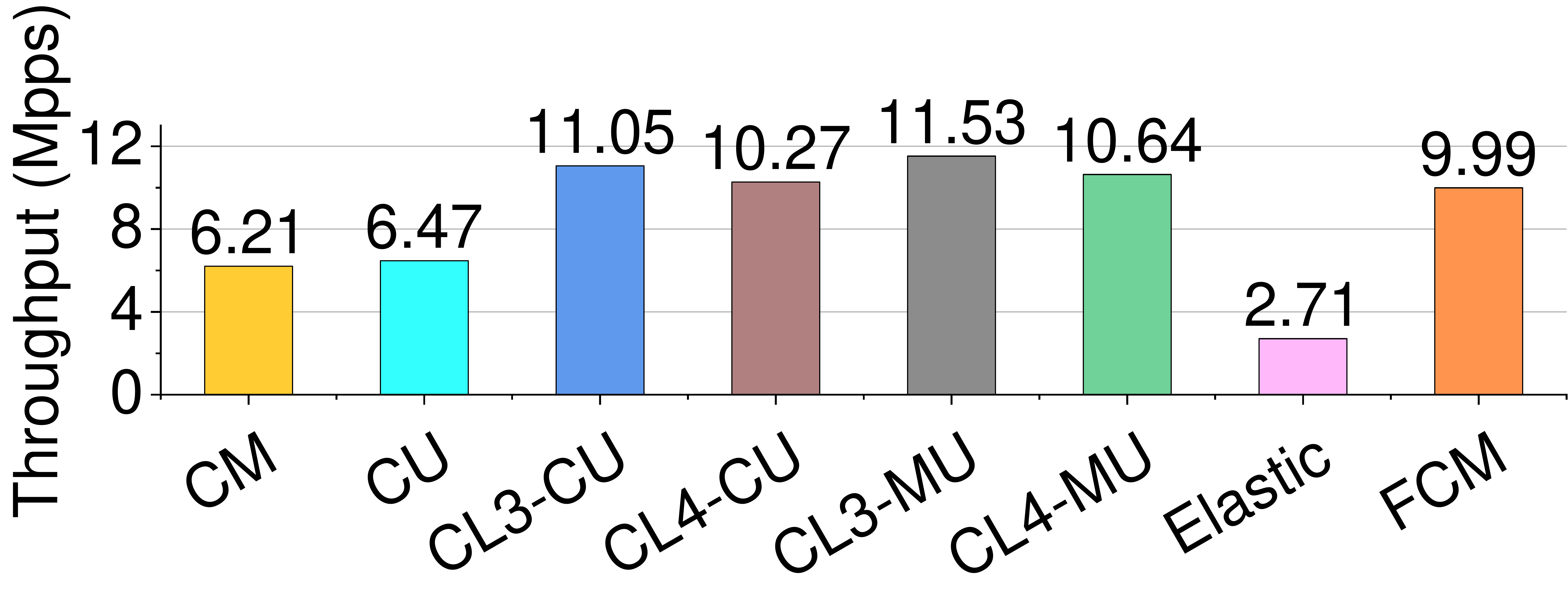}
    \caption{
		     Throughput: packet encoding capacity in a CPU environment. Mpps: million packets per second.
		}~\label{fig:throughput}
\end{figure}

In the following, we evaluate \ours{}'s performance with actual network traffic measurement tasks, including flow size estimation, cardinality, flow size distribution, entropy, heavy hitter, heavy changer, and throughput. 
Fig.~\ref{fig:cpu_exp_1} shows a comparison of switch ASIC-friendly \ours{} (CL-MU), \cmin{} (CM), Elastic sketch, and FCM sketch in terms of their accuracy.
Fig.~\ref{fig:throughput} provides a comparison analysis of the throughput.

\BfPara{Flow Size Estimation}
Flow size estimation entails a per-flow packet counting of all individual flows in an epoch. As expected, \clmu{} with a four-layer setting (CL4-MU) achieves the best accuracy, as shown in Fig.~\ref{fig:cpu_exp_1}. With the same layer setting, \clmu{} is much more accurate than the standard \cmin{}. Moreover, \clmu{} outperforms the Elastic and FCM sketches in general. We note that the results are in line with our theoretical and experimental analysis.

\BfPara{Cardinality}
In cardinality estimation, we count the number of distinct flows in a network trace. \ours{} uses its own data structure and the linear counting (LC) theory to estimate the cardinality~\cite{Whang:1990}. To do so, we assume the lowest layer of \ours{} to be the LC's bit array. Then, the cardinality is calculated following the LC's equation, $\hat{n}=-s\cdot \ln(V)$, where $s$ is the number of counters and $V$ is the fraction of  the 
empty (zero) counters in the array. 
Since \ours{}'s number of counters at the lowest layer is 2x-16x larger than CM, \ours{} is saturated relatively slower. Given LC's performance is guaranteed only until the number of empty counters is less than 30\%, \ours{} can estimate the cardinality with smaller memory. The results in Fig.~\ref{fig:cpu_exp_1}(b) confirm that CM cannot provide a valid estimation with small memory. 
\ours{} with a three-layer setting (CL3-MU) achieves a similar accuracy to the Elastic sketch and outperforms the FCM sketch noticeably with small memory. Moreover, \ours{} achieves the best accuracy when we extend the number of layers from three to four (\ie CL4-MU).

\BfPara{Flow Size Distribution}
Different from Elastic sketch that employs the computationally heavy MRAC~\cite{kumar2004data}, we leverage the distribution array that is recorded on-the-fly to estimate the distribution of the mouse flows. For elephant flows, we query our sketch with the label table used in the heavy hitter and heavy changer detection. 
In Fig.~\ref{fig:cpu_exp_1}(c), we show the accuracy of the flow size distribution among all sketches. As shown, \ours{} outperformed CM always, achieving WMRE of 0.45 and 0.34 on average when $d$=3 and 4, respectively. However, \ours{} was less accurate than Elastic and FCM, which had WMRE of 0.17 and 0.42 on average. We note that Elastic and FCM rely on MRAC for the estimation, which requires a very large overhead in decoding and has to be done off-line (\ie $\mathcal{O}(\#counters^3)$). \ours{}'s flow size distribution estimation can be done online while achieving a sufficient accuracy using a lightweight array and a label table, and no additional computations (\ie $\mathcal{O}(1)$).

\BfPara{Entropy}
For entropy measurement, we used a 255 $\times$ 32-bit counter array to record the distribution of mouse flows (\eg 1$\sim$255) on-the-fly, where the index is the size of a flow and the counter value is the frequency. Subsequently, elephant flows will be involved by querying our sketch with the label table.
Finally, the entropy is calculated by leveraging the distribution array by $-\sum{i*\frac{n_i}{N}*log\frac{n_i}{N}}$, where $N$ is the total number of flows and $n_i$ is the number of flows with $i$ packets~\cite{yang2018elastic}.
In Fig.~\ref{fig:cpu_exp_1}(d), \ours{} is shown to outperforms Elastic sketch in all memory size and with all configurations, with an average RE of 0.0082 and 0.0058 when s$d=3$ and $d=4$, respectively. With memory size from 0.2 MB to 0.6 MB, \ours{} is 1.21-7.84 times more accurate than FCM.
When using 1.0 MB memory, FCM (0.0009) is better than CL3-MU (0.0023) and CL4-MU (0.0039). We note that the accuracy of FCM is obtained by MRAC, thus it requires more resources for this performance.

\BfPara{Heavy Hitter}
For heavy hitter and changer detection, we assigned 0.1 MB of memory for a label table, which is responsible for storing large flows.
In our experiments, a flow that contributes more than 0.1\% (2.5K$\sim$2.7K) of the total number of packets in each epoch is defined as a heavy hitter and stored in our label table. 
We note that the recall rate of \ours{}'s heavy hitter detection is always 1.0 because our sketch will never underestimate a flow. For precision rate, \ours{} also shows a good performance, since the mouse flows will not flood elephant flows, thanks to our minimum update strategy. Except for CM and FCM with 0.2 MB memory (F1 Score of 0.978 and 0.974), we found that all sketches achieved a fair performance.

\BfPara{Heavy Changer}
The heavy changers are flows whose size changes (increases or decreases) significantly over two adjacent epochs. To detect them, all sketches consider flows recorded in the heavy hitter table based on a threshold. Here, we use 0.1\% of the total changes over two adjacent epochs as a threshold. 
As we can find in Fig.~\ref{fig:cpu_exp_1}(c), CL4-MU and FCM with 0.2 MB memory posts an F1 score of 0.970 and 0.950, which are slightly lower than other schemes. The reason why CL4-MU is less accurate compared to CL3-MU is because CL4-MU assigns more counters at the lower layers and fewer counters at the upper layers (\ie mouse flow friendly), thereby the heavy changer detection is less accurate with small memory. However, CL4-MU is able to recover F1 score as high as the other sketches with larger memory.

\BfPara{Throughput}
We measure sketch's throughput in the CPU environment to demonstrate the encoding complexity of different schemes. As shown in Fig.~\ref{fig:throughput}, \ours{}'s throughput is roughly twice higher than the other sketches. CL3-MU and CL4-MU achieve the highest throughput of 11.53 Mpps and 10.64 Mpps, respectively. The optimal CL3-CU and CL4-CU are slightly lower than CL-MU (\ie 11.05 Mpps and 10.27 Mpps) due to the global minimum-based update algorithm, which requires additional memory access.  FCM sketch shows better throughput (9.99 Mpps) than CM, CU, and Elastic sketches and comparable to CL4-CU (10.27 Mpps). Elastic sketch posted the lowest throughput of 2.71 Mpps. However, we note that Elastic sketch was designed to estimate the source IP-based flows (\ie 32-bit label). Therefore, the discrepancy between Elastic's throughput and the results posted in the original work are due to operating with the 5-tuple label (\ie 104 bits), which cannot be accelerated by Intel's Streaming SIMD Extensions (SSE) technologies~\cite{SIMD}.

\section{Data Plane Implementation and Evaluation}\label{sec:p4_impl}

In this section, we present the implementation details and evaluations of our ASIC-friendly \ours{} (\ie \clmu{} and \clcm{}). Particularly, we first describe the hardware implementation of \ours{} sketch in the programmable Tofino switch. Moreover, we compare \ours{} with the standard and state-of-the-art sketches in terms of resource usage, used stage, and the packet processing latency. Finally, we verify \ours{}'s performance using our prototype.

\subsection{Implementation}
We implemented a prototype of \clmu{} sketch among applications in the data plane and control plane of Tofino Wedge-100 switch~\cite{tofino_wedge} using the P4 language. For the data plane implementation, we added 115 lines of P4 code: 93 lines for the data structure and function declaration and 22 lines for the function execution. The data plane program is compiled with Barefoot Software Development Environment 9.0.0~\cite{tofino_wedge}.
In the meantime, the network traffic measurement applications like flow size estimation, cardinality, flow size distribution, and entropy were implemented in the control plane using Bfrt Python~\cite{hauser2021survey}. The communication between the data and control planes is realized through the P4Runtime API~\cite{P4Runtime_API}.

\BfPara{(1) Infeasibility of the optimal CL-CU in Data Plane}
Our optimal CL-CU sketch's update algorithm is composed of a two-step operation, namely (S1) reading counters of a flow across all layers and (S2) updating the minimum counter among them.
Hardware-wise (\ie switch ASIC), step (S2) requires revisiting the register array (\ie layer) that hosts the minimum counter for updating.
Unfortunately, such double-access of the same memory register is not allowed after the processing stage jumping to the next memory region (layer) during the step (S1) due to the pipeline design~\cite{zhang2020poseidon, Barefoot_Doc}.
One can resolve the issue by leveraging packet re-circulation design. However, that design choice will hurt the bandwidth of the switch, which may not be acceptable for use in a high-speed environment~\cite{jin2018netchain, jin2017netcache, sivaraman2017heavy}.

\BfPara{(2) ASIC-friendly CL-MU Implementation}
To resolve the aforementioned issue, we designed the ASIC-friendly CL-MU, which is an approximate version of CL-CU, to embrace the switch's pipeline. 
For data structure, each layer's counter array is implemented by register arrays and hosted by a single stage of the ASIC's packet processing pipeline. Moreover, the interaction operations between layers are integrated into the register action logic. Simply put, each layer was deployed sequentially in different match+action units (MAUs) over the pipeline.

\begin{figure*}[t]
    
    \centering
    \includegraphics[width=0.85\textwidth]{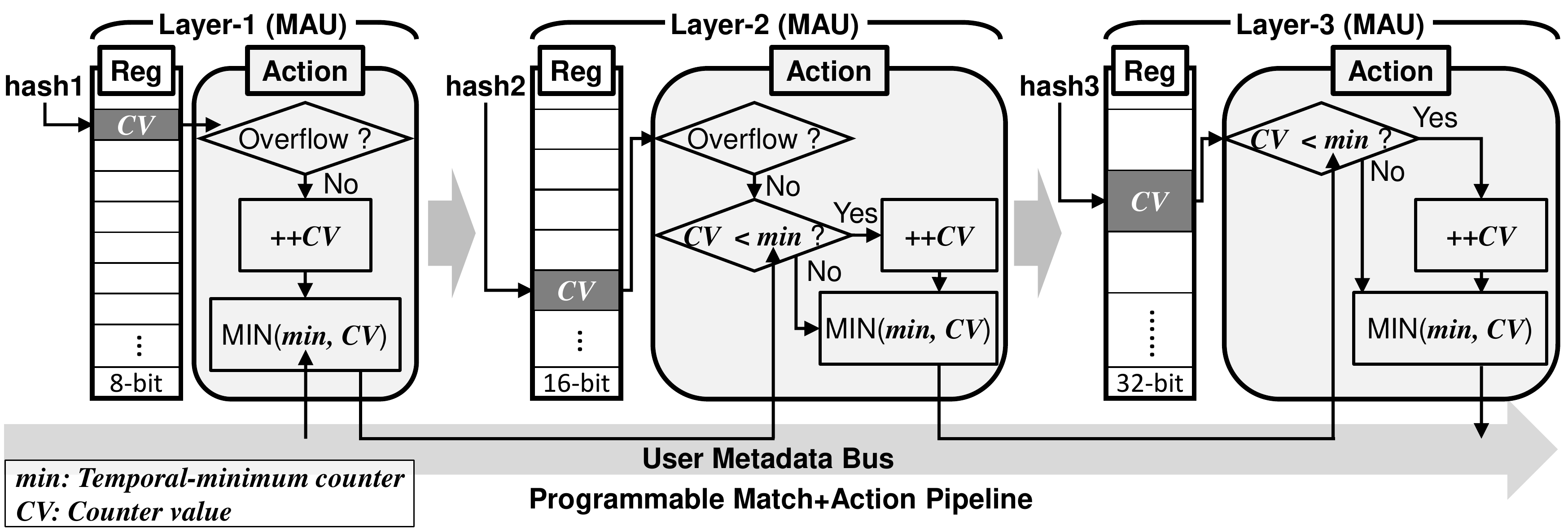}
    \caption{
		     Pipeline design of \ours{}: encoding/decoding logic of \clcu{} with the three-layer setting.
		}~\label{fig:mu_update_p4}
\end{figure*}

\begin{figure}[t]
    \centering
    \includegraphics[width=0.48\textwidth]{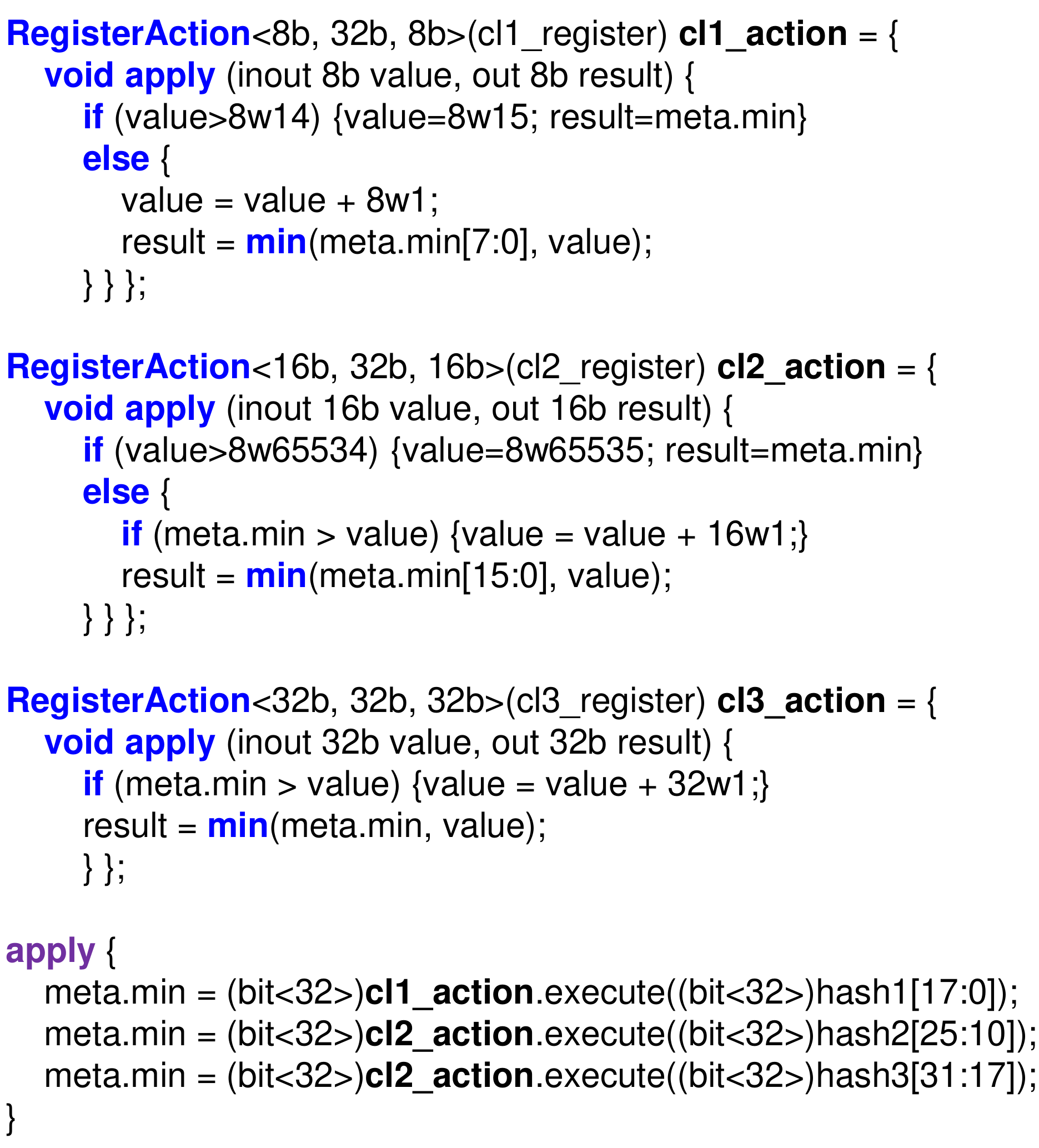}
    \caption{
	     Data plane implementation of \ours{} sketch in P4.
	}~\label{fig:p4_code}
\end{figure}

Fig.~\ref{fig:mu_update_p4} presents the logic of CL3-MU in the data plane.
Three CRC32-based hash functions ~\cite{peterson1972error} were used to locate the counters for lookup and update of counter values (\textit{CVs}) of a flow $f$ in each layer.
During the first stage of layer update, it checks if the current flow's counter reaches its counting capacity. If the counter is overflowed, we skip the counter update of the current stage (layer) and go to the next stage for the next layer's counter update.
Otherwise, we update the counter value (\textit{CV}) by increasing it by one. In the meantime, we read the pre-initialized minimum value (\textit{min}) from the user metadata bus through the gateway and update  the (\textit{min}) with \textit{CV}. We note that all these actions are performed with the arithmetic logical units (ALUs) that are associated with the current stage's MAU.
The updated \textit{min} will be stored into metadata and passed to a subsequent stage (MAU) for information sharing.
The similar operations are repeated at the next stage (\ie layer-2) with an additional minimum value check between the flow's layer-2 counter $CV$ and the $min$ from the previous stage (\ie layer-1). $CV$ is updated only if it is the minimum counter upon the current stage. Then, $min$ in the user metadata is updated for the next layer's operation. At the last stage, we assume the counter cannot be overflowed due to a sufficiently large register (\ie 32-bit). Therefore, we repeat the rest of the processes of layer-2.
The P4 code snippet of the essential logic of \ours{} is shown in Fig.~\ref{fig:p4_code}.

\BfPara{(3) Control Plane}
The application's estimation functions are implemented using Bfrt Python in the control plane.
To retrieve statistics from the data plane, the control plane application of \ours{} fetches register values of each layer, flow size distribution array, and heavy hitter table through P4Runtime API~\cite{P4Runtime_API} for post-hoc analysis.
Remarkably, due to the data plane deployment of the distribution array and heavy hitter table, and the online decoding capacity of \ours{}, our prototype is capable of performing all tasks without a powerful server.

\begin{figure}[t]
    \centering
	\includegraphics[width=0.32\textwidth]{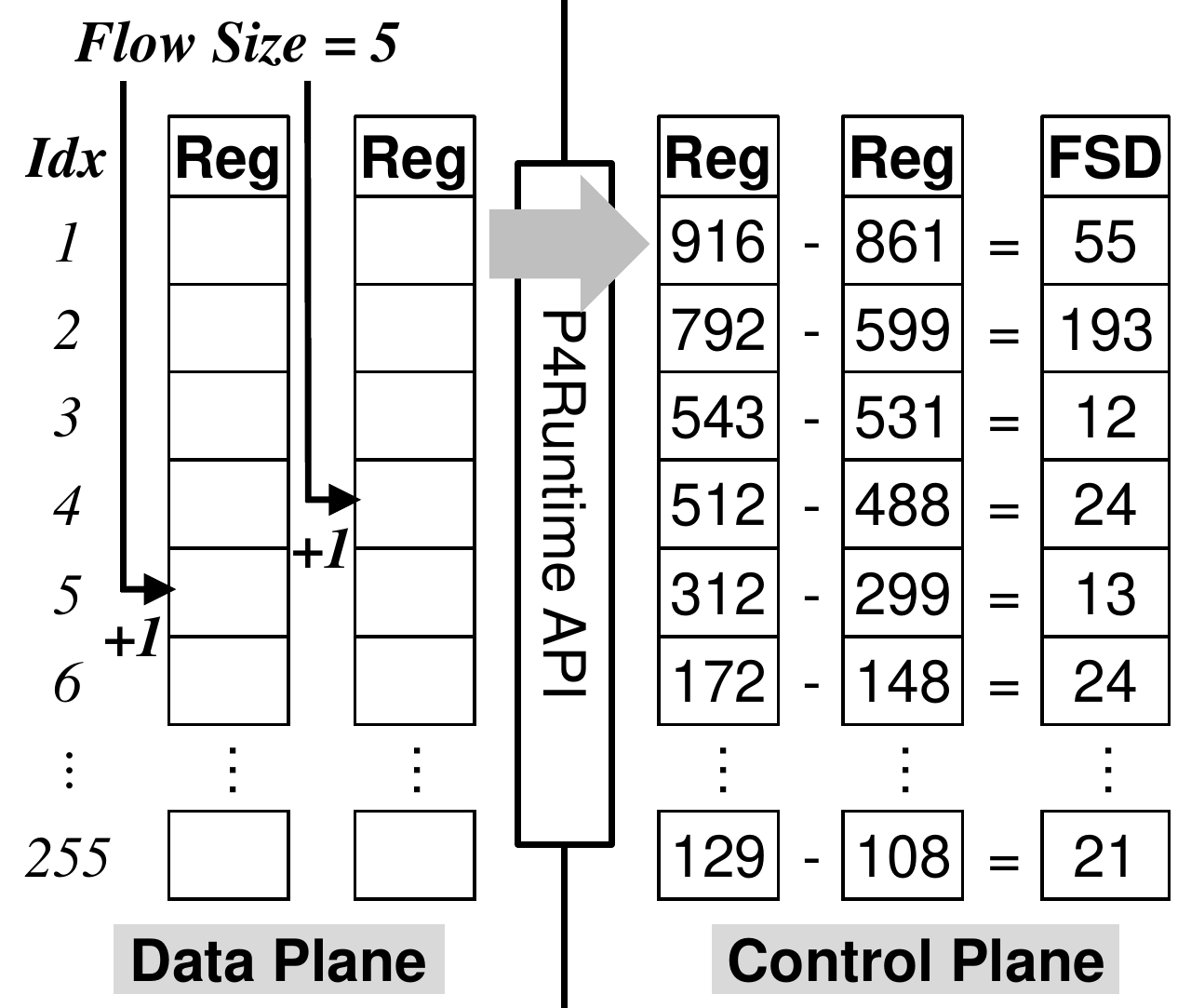}
    \caption{
	     The data structure and logic of our data plane flow size distribution array.
	}~\label{fig:dist_arr_p4}
\end{figure}

\BfPara{(4) Flow Size Distribution Array}
The flow size distribution array is a counter array, which measures the frequency of different sized flows while encoding and decoding flows. This counter array uses the index as a flow size and the counter value as the flow's frequency. Due to the massive range of the flow size, we limit the flow size (\ie index) to 255 and measure the distribution of mouse flows only. When a flow's size is updated (say, 4 to 5), we increase the fifth counter by ``1'' and decrease the fourth counter by ``1'' to adjust the distribution. Unfortunately, due to the memory double-access constraint of the data plane, we cannot access two counters of a single array in a packet processing cycle.  To resolve the issue, we implemented the distribution array with two register arrays, as shown in Fig.~\ref{fig:dist_arr_p4}. After the encoding and decoding of a flow, one register array increases the frequency of the updated flow size, and the other register array also increases the frequency but for the flow size before updating. Eventually, we obtain the desired flow size distribution by performing counter-wise subtraction between the first and the second array.  We note that the distribution arrays reside in the data plane to record flow sizes on the fly. For decoding the flow size distribution, we retrieve the register arrays from the data plane to the control plane via the P4Runtime API~\cite{P4Runtime_API} for the estimation.

\subsection{Hardware Evaluation}
In the following, we compare \ours{} with data plane-deployable solutions in terms of resource usage, stage, latency, and accuracy. 
We note that \ours{} P4 implementation was successfully compiled and executed in our switch, allowing \ours{} to process packets at line-rate~\cite{miao2017silkroad, yang2018elastic, jin2018netchain}.

\BfPara{(1) Settings} 
In the hardware evaluation, we fixed the memory size of sketches as 0.6 MB. The testbed consists of one Tofino programmable switch with 32x100 Gbps ports and one commodity client PC with AMD Ryzen 5 2400 G 8-core CPU, 16 GB DRAM, and Intel 40 Gbps network adapter, all connected by 40 Gbps QSFP+ cables.
We used a network packet crafting library called libtins~\cite{libtins_pkt_gen} to replay the CAIDA traffic between the client PC and our programmable switch. In the control plane, four applications were executed for comparisons among sketches.

\begin{figure*}[t]
    \centering
    \subfigure[Resource Usage]{\includegraphics[width=0.24\textwidth]{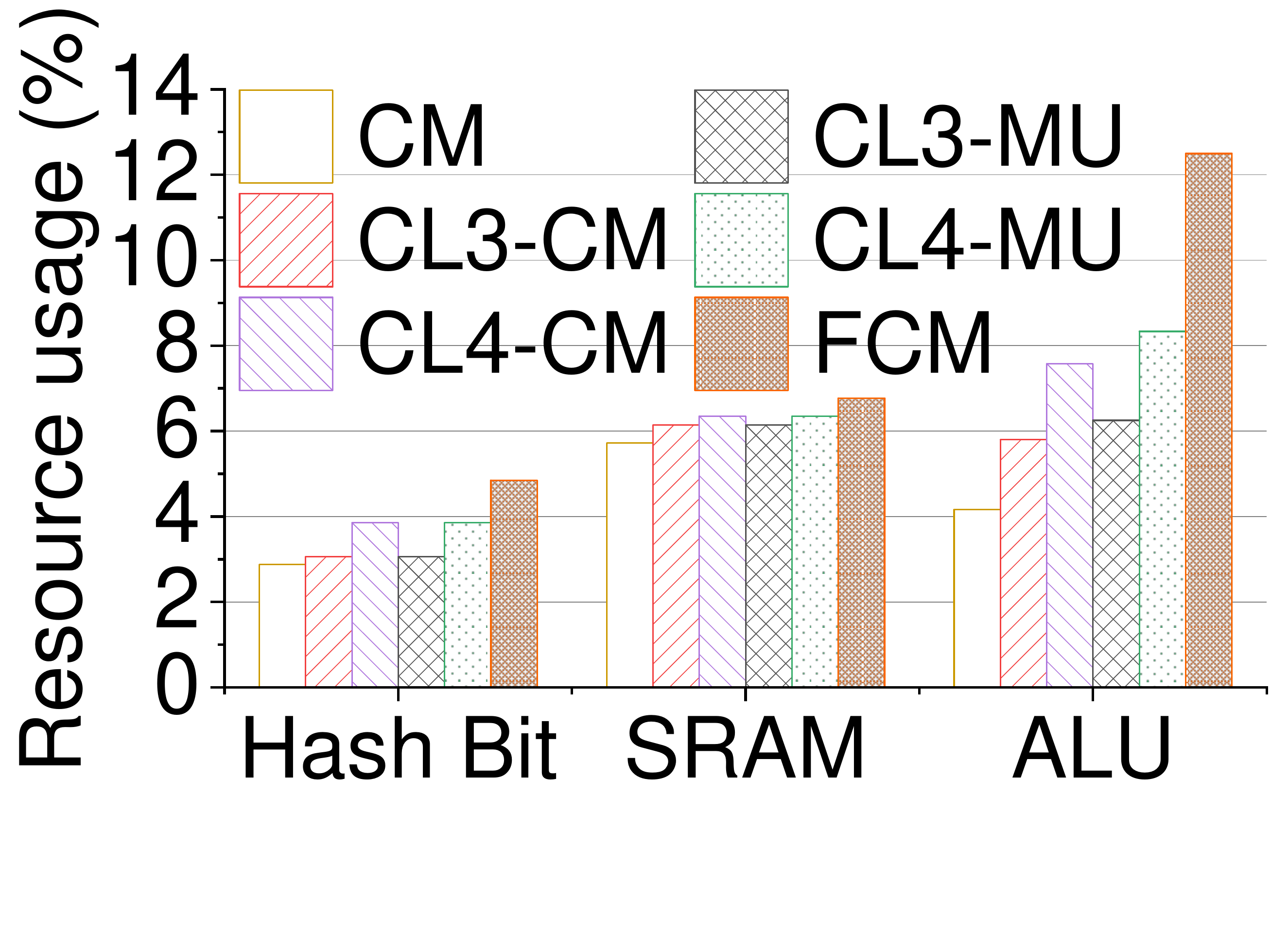}}
    \hspace{6mm}
    \subfigure[Stage]{\includegraphics[width=0.24\textwidth]{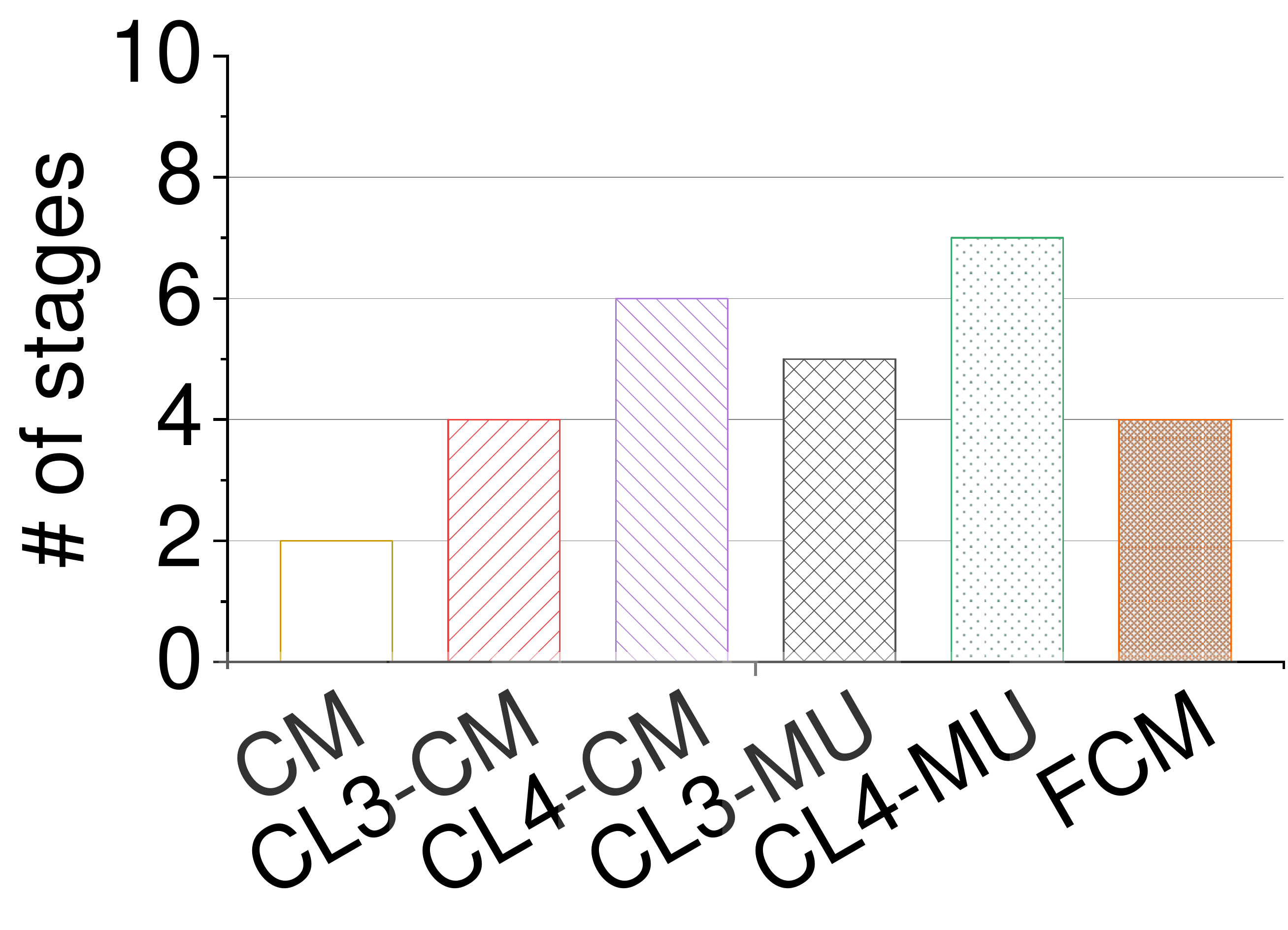}}
    \hspace{6mm}
    \subfigure[Normalized Latency]{\includegraphics[width=0.28\textwidth]{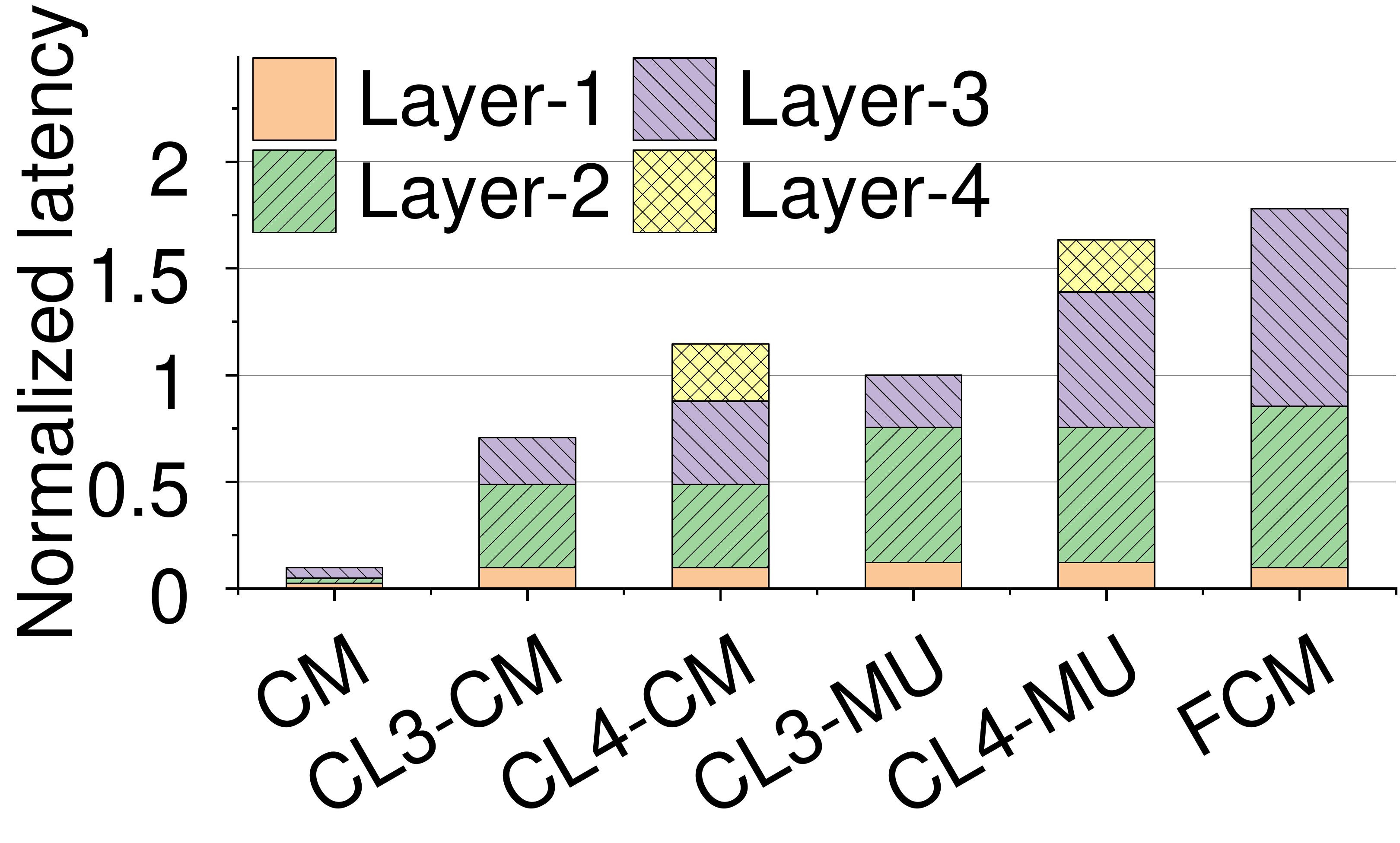}}
    \caption{Resource consumption and latency:
            (a) and (b) show a comparison of the resource usage and used stage in the data plane for \ours{}, \cmin{} and FCM sketches. (c) demonstrates the normalized packet processing latency.
		}~\label{fig:p4_eval_01}
\end{figure*}

\BfPara{(2) Resources Usage}
We first discuss the resource usage of \ours{}, \cmin{}, and FCM sketches. Elastic sketch's result is not included since their P4 code is not publicly available. 
As shown in Fig.~\ref{fig:p4_eval_01}(a), \cmin{} sketch consumes 2.88\% of hash bit, 5.72\% of SRAM, and 4.16\% of ALU resources, which are the lowest among the compared sketches.
CL3 requires slightly higher hash bits and SRAM compared to \cmin{} (\ie 3.06\% of hash bit and 6.14\% of SRAM),  followed by CL4's 3.86\% and 6.35\%. CL-CM requires fewer ALU resources than CL-MU due to the simpler update algorithm.
Finally, FCM sketch requires 4.97\% of hash bit, 7.29\% of SRAM, and 16.67\% of ALU, which are the highest.
The high cost of FCM sketch is caused by its multi-tree design, which repeats the same process multiple times for independent sketches with different hash functions. Therefore, FCM's high ALU utilization indicates its complexity, which hurts the packet encoding latency of the data plane (see Fig.~\ref{fig:p4_eval_01}(c)).

\BfPara{(3) Stage and Packet Processing Latency}
As shown in Fig.~\ref{fig:p4_eval_01}(b), our advanced update algorithm sacrifices several stages for payment. Moreover, the cost increases as the layers increase. However, CL3-CM, with the same stage numbers as FCM, is comparable in terms of accuracy as shown in Table~\ref{table:Acc}, while CL3-CM's latency is only 42.9\% of FCM's, as shown in Fig.~\ref{fig:p4_eval_01}(c). With one stage overhead, CL3-MU outperforms FCM except FSD, but CL3-MU's latency is only 57.1\% of FCM's. That is, the stage overhead does not necessarily mean an additional latency of our sketch.  
As shown in Fig.~\ref{fig:p4_eval_01}(c), \ours{} sketch's the middle layers (\ie layer-2 in CL3, and layer-2 and 3 in CL4) contribute the most to the total latency (66.07\% on average) followed by the top layer's  23.44\% and the bottom layer's 10.49\%. However, the overall latency is lower than that of FCM sketch, which contributes 52.05\%, 42.46\%, and 5.49\% of latency by layer-3, layer-2, and layer-1, respectively.
This result suggests that appropriate separation of the logic operations into more stages reduces the processing latency, which is a better design than packing complex operations into fewer stages. Moreover, dye to the resource management function of the Barefoot compiler, the idle resources at each stage can be optimized and assigned for other data plane functions for maximizing the resource utilization rate.


\begin{table*}[t]
    
	\renewcommand{\arraystretch}{1.0}
	\centering
	\caption{Comparison: Hardware-based network traffic measurement applications with 0.6 MB memory space. Lower values mean better accuracy.}
	\label{table:Acc}
	{\footnotesize
	\scalebox{1.2}{
	\centering
	\begin{tabular}{l|r|r|r|r|r|r}
		\Xhline{2\arrayrulewidth}
		Sketch & CM & CL3-CM & CL4-CM & CL3-MU & CL4-MU & FCM\\
		\Xhline{2\arrayrulewidth}
		Flow Size Estimation (ARE)	&	7.181	&	1.641	&	1.021	&	1.171	&	\bf{0.601}	&	1.591	\\
		Cardinality (RE)	&	0.007	&	0.003	&	0.003	&	0.002	&	\bf{0.001}	&	0.002	\\
		Flow Size Distribution (WMRE)	&	1.159	&	0.399	&	0.411	&	0.294	&	0.312	&	\bf{0.291}	\\
		Entropy (RE)	&	0.018	&	0.009	&	0.007	&	\bf{0.005}	&	\bf{0.005}	&	0.006	\\
		\Xhline{2\arrayrulewidth}
	\end{tabular}}}
\end{table*}

\BfPara{(4) Accuracy Comparison}\label{sec:p4_eval_acc}
Table~\ref{table:Acc} shows accuracy comparisons among \cmin{}, \clcm{}, \clmu{} and FCM sketches by varying applications, namely flow size estimation (ARE), cardinality (RE), flow size distribution (WMRE), and entropy (RE).
As shown, the standard \cmin{} performs the worst among the compared sketches. On the other hand, CL4-MU provides the best overall performance. Remarkably, CL3-MU also outperforms FCM in flow size and entropy estimations and achieves a comparable accuracy in cardinality and flow size distribution estimations. 
Moreover, the lightweight CL-CM with a four-layer setting (CL4-CM) provides a reasonable accuracy compared to the FCM sketch. Such results are in line with our observations in the CPU experiments. Considering the resource consumption and latency performance, we conclude that our \ours{} with minimum update algorithm is superior to the state-of-the-art FCM sketch. Moreover, \ours{} sketch family (CL3-CM, CL3-MU, CL4-CM, CL4-MU) provides more options to users for balancing the needs between latency and accuracy.

\section{Related Work}\label{sec:relatedwork}
\cmin{} (CM) sketch~\cite{cormode2005improved} is a probabilistic data structure capable of representing a high-dimensional vector and answering queries on this vector. CM sketch can measure the spectral density of the high volume of data using a small amount of memory. Therefore, it is adopted in various fields, such as network monitoring. However, an identical size of $w$ counters and an indiscriminate encoding/decoding of \cmin{} lead to poor scalability and memory efficiency, causing estimation degradation. \cmcu{}~\cite{goyal2012sketch} mitigated \cmin{}'s scalability problem by updating a minimum number of counters to saturate more slowly. However, there are still two issues: low memory utilization and infeasibility in the data plane in the P4 switch.

To date, a large body of measurement systems are proposed, including  OpenSketch~\cite{yu2013software}, Dream~\cite{MoshrefYGV14DREAM}, Pingmesh~\cite{GuoYXDHMLWPCLK15Pingmesh}, UnivMon~\cite{liu2016one}, Trumpet~\cite{MoshrefYGV16Trumpet}, FlowRadar~\cite{li2016flowradar}, SketchVisor~\cite{huang2017sketchvisor},  Marple~\cite{NarayanaSNGAAJK17Marple}, Elastic Sketch~\cite{yang2018elastic}, and FCM sketch~\cite{song2020FCM}. Among them, FlowRadar, UnivMon, Elastic Sketch, FCM sketch are generic and are shown feasible for a programmable switch. 
FlowRadar~\cite{li2016flowradar} stores all flow IDs and flow sizes in a Bloom Filter and an Invertible Bloom Lookup Table through XOR operations. The memory usage for FlowRadar is proportional to the number of flows we need to store.
UnivMon~\cite{liu2016one} uses universal streaming, a top-k detection sketch that guarantees the accuracy for network traffic monitoring. While generic with good performance, UnivMon does not take into account the various traffic characteristics. 
Elastic Sketch~\cite{yang2018elastic} provides a separation technique named Ostracism to keep the elephant flows from mouse flows and is composed of two parts: the heavy part recording elephant flows and the light recording mouse flows. The heavy part is a hash table, where the elephant flows are stored and evicted to the light part with a voting method for flow size. The light part is a \cmin{} sketch tracking the mouse flows with $d$ hash functions. With both parts, Elastic Sketch works well in terms of accuracy for various tasks.
FCM sketch~\cite{song2020FCM} uses a similar idea of the pyramid counter~\cite{YangZJCL17Pyramid} for systematic filtering of flows depending on flow sizes. It also extends the idea with a multi-tree design for reducing hash collision.
Among them, FCM sketch is the most related work to ours.

\section{Conclusion}\label{sec:conclusion}
Low latency and high accuracy network measurement is still paramount and challenging. \cmin{} sketch is one of the best candidate solutions owing to its accuracy and speed, but its data structure and the indiscriminate update strategy lead to poor scalability, memory waste, and estimation degradation. To address these issues, we presented a new sketch called \ours{}, which leverages the strategy of {\em Split Counter} and {\em Minimum Update}. To show its feasibility, we conducted extensive analysis, theoretically and experimentally. Moreover, we designed an ASIC-friendly version of \ours{}, deployed it in a programmable switch, and performed a comprehensive evaluation by comparing it with state-of-the-art sketches. Our results prove that \ours{} has a well-balanced performance and can be used in many applications, standalone or as a supporting tool.

\bibliographystyle{IEEEtran}
\bibliography{reference}

\begin{thebibliography}{10}
\providecommand{\url}[1]{#1}
\csname url@samestyle\endcsname
\providecommand{\newblock}{\relax}
\providecommand{\bibinfo}[2]{#2}
\providecommand{\BIBentrySTDinterwordspacing}{\spaceskip=0pt\relax}
\providecommand{\BIBentryALTinterwordstretchfactor}{4}
\providecommand{\BIBentryALTinterwordspacing}{\spaceskip=\fontdimen2\font plus
\BIBentryALTinterwordstretchfactor\fontdimen3\font minus
  \fontdimen4\font\relax}
\providecommand{\BIBforeignlanguage}[2]{{%
\expandafter\ifx\csname l@#1\endcsname\relax
\typeout{** WARNING: IEEEtran.bst: No hyphenation pattern has been}%
\typeout{** loaded for the language `#1'. Using the pattern for}%
\typeout{** the default language instead.}%
\else
\language=\csname l@#1\endcsname
\fi
#2}}
\providecommand{\BIBdecl}{\relax}
\BIBdecl

\bibitem{BGPstat}
\BIBentryALTinterwordspacing
\emph{BGP table data}, Accessed June 20, 2020. [Online]. Available:
  \url{https://bgp.potaroo.net/}
\BIBentrySTDinterwordspacing

\bibitem{YHS21self}
Y.~Du, H.~Huang, Y.-E. Sun, S.~Chen, and G.~Gao, ``Self-adaptive sampling for
  network traffic measurement,'' in \emph{IEEE INFOCOM 2021}, 2021.

\bibitem{jang2020sketchflow}
R.~Jang, D.~Min, S.~Moon, D.~Mohaisen, and D.~Nyang, ``Sketchflow: Per-flow
  systematic sampling using sketch saturation event,'' in \emph{IEEE INFOCOM
  2020}.\hskip 1em plus 0.5em minus 0.4em\relax {IEEE}, 2020, pp. 1339--1348.

\bibitem{kumar2006sketch}
A.~Kumar and J.~J. Xu, ``Sketch guided sampling - using on-line estimates of
  flow size for adaptive data collection,'' in \emph{INFOCOM 2006}.\hskip 1em
  plus 0.5em minus 0.4em\relax {IEEE}, 2006.

\bibitem{liu2019nitrosketch}
Z.~Liu, R.~Ben-Basat, G.~Einziger, Y.~Kassner, V.~Braverman, R.~Friedman, and
  V.~Sekar, ``Nitrosketch: Robust and general sketch-based monitoring in
  software switches,'' in \emph{ACM SIGCOMM 2019}, 2019, pp. 334--350.

\bibitem{Morris78a}
R.~H.~M. Sr., ``Counting large numbers of events in small registers,''
  \emph{Commun. {ACM}}, vol.~21, no.~10, pp. 840--842, 1978.

\bibitem{flajolet1985probabilistic}
P.~Flajolet and G.~N. Martin, ``Probabilistic counting algorithms for data base
  applications,'' \emph{Journal of computer and system sciences}, vol.~31,
  no.~2, pp. 182--209, 1985.

\bibitem{cormode2005improved}
G.~Cormode and S.~Muthukrishnan, ``An improved data stream summary: the
  count-min sketch and its applications,'' \emph{Journal of Algorithms},
  vol.~55, no.~1, pp. 58--75, 2005.

\bibitem{kumar2004data}
A.~Kumar, M.~Sung, J.~Xu, and J.~Wang, ``Data streaming algorithms for
  efficient and accurate estimation of flow size distribution,'' \emph{ACM
  SIGMETRICS PER}, vol.~32, no.~1, pp. 177--188, 2004.

\bibitem{estan2002new}
C.~Estan and G.~Varghese, ``New directions in traffic measurement and
  accounting,'' in \emph{ACM SIGCOMM 2002}, 2002, pp. 323--336.

\bibitem{Whang:1990}
K.~Whang, B.~T.~V. Zanden, and H.~M. Taylor, ``A linear-time probabilistic
  counting algorithm for database applications,'' \emph{{ACM} Trans. Database
  Syst.}, vol.~15, no.~2, pp. 208--229, 1990.

\bibitem{nyang16}
D.~Nyang and D.~Shin, ``Recyclable counter with confinement for real-time
  per-flow measurement,'' \emph{{IEEE/ACM} Trans. Netw.}, vol.~24, no.~5, pp.
  3191--3203, 2016.

\bibitem{zhou2018cold}
Y.~Zhou, T.~Yang, J.~Jiang, B.~Cui, M.~Yu, X.~Li, and S.~Uhlig, ``Cold filter:
  A meta-framework for faster and more accurate stream processing,'' in
  \emph{SIGMOD/PODS 2018}, 2018, pp. 741--756.

\bibitem{YangZJCL17Pyramid}
T.~Yang, Y.~Zhou, H.~Jin, S.~Chen, and X.~Li, ``Pyramid sketch: a sketch
  framework for frequency estimation of data streams,'' \emph{Proc. {VLDB}
  Endow.}, vol.~10, no.~11, pp. 1442--1453, 2017.

\bibitem{LuMPDK08CounterBraids}
Y.~Lu, A.~Montanari, B.~Prabhakar, S.~Dharmapurikar, and A.~Kabbani, ``Counter
  braids: a novel counter architecture for per-flow measurement,'' in
  \emph{Proc. {ACM} {SIGMETRICS} 2008}, Z.~Liu, V.~Misra, and P.~J. Shenoy,
  Eds.

\bibitem{song2020FCM}
C.~H. Song, P.~G. Kannan, B.~K.~H. Low, and M.~C. Chan, ``Fcm-sketch: Generic
  network measurements with data plane support,'' in \emph{CoNext 2020}.\hskip
  1em plus 0.5em minus 0.4em\relax ACM, 2020, pp. 78--92.

\bibitem{huang2017sketchvisor}
Q.~Huang, X.~Jin, P.~P.~C. Lee, R.~Li, L.~Tang, Y.~Chen, and G.~Zhang,
  ``Sketchvisor: Robust network measurement for software packet processing,''
  in \emph{ACM SIGCOMM 2017}.\hskip 1em plus 0.5em minus 0.4em\relax {ACM},
  2017, pp. 113--126.

\bibitem{yang2018elastic}
T.~Yang, J.~Jiang, P.~Liu, Q.~Huang, J.~Gong, Y.~Zhou, R.~Miao, X.~Li, and
  S.~Uhlig, ``Elastic sketch: Adaptive and fast network-wide measurements,'' in
  \emph{ACM SIGCOMM 2018}, 2018, pp. 561--575.

\bibitem{tirmazi2020cheetah}
M.~Tirmazi, R.~B. Basat, J.~Gao, and M.~Yu, ``Cheetah: Accelerating database
  queries with switch pruning,'' in \emph{ACM SIGMOD 2020}, D.~Maier,
  R.~Pottinger, A.~Doan, W.~Tan, A.~Alawini, and H.~Q. Ngo, Eds.\hskip 1em plus
  0.5em minus 0.4em\relax {ACM}, 2020, pp. 2407--2422.

\bibitem{zhang2020poseidon}
M.~Zhang, G.~Li, S.~Wang, C.~Liu, A.~Chen, H.~Hu, G.~Gu, Q.~Li, M.~Xu, and
  J.~Wu, ``Poseidon: Mitigating volumetric ddos attacks with programmable
  switches,'' in \emph{NDSS 2020}.\hskip 1em plus 0.5em minus 0.4em\relax The
  Internet Society, 2020.

\bibitem{goyal2011approximate}
A.~Goyal and H.~D. III, ``Approximate scalable bounded space sketch for large
  data {NLP},'' in \emph{EMNLP 2011}.\hskip 1em plus 0.5em minus 0.4em\relax
  {ACL}, 2011, pp. 250--261.

\bibitem{goswami2018buffered}
M.~Goswami, D.~Medjedovic, E.~Mekic, and P.~Pandey, ``Buffered count-min sketch
  on ssd: Theory and experiments,'' \emph{arXiv preprint arXiv:1804.10673},
  2018.

\bibitem{deng2007new}
F.~Deng and D.~Rafiei, ``New estimation algorithms for streaming data:
  Count-min can do more,'' \emph{Webdocs. Cs. Ualberta. Ca}, 2007.

\bibitem{goyal2012sketch}
A.~Goyal, H.~Daum{\'e}~III, and G.~Cormode, ``Sketch algorithms for estimating
  point queries in nlp,'' in \emph{In Joint Conference on EMNLP/CoNLL 2012},
  2012, pp. 1093--1103.

\bibitem{yu2013software}
M.~Yu, L.~Jose, and R.~Miao, ``Software defined traffic measurement with
  opensketch,'' in \emph{NSDI 2013}, N.~Feamster and J.~C. Mogul, Eds.\hskip
  1em plus 0.5em minus 0.4em\relax {USENIX} Association, 2013, pp. 29--42.

\bibitem{pitel2015cmlog}
G.~Pitel and G.~Fouquier, ``Count-min-log sketch: Approximately counting with
  approximate counters,'' \emph{International Symposium on Web Algorithms},
  2015.

\bibitem{pitel2016cmtree}
\BIBentryALTinterwordspacing
G.~Pitel, G.~Fouquier, E.~Marchand, and A.~Mouhamadsultane, ``Count-min tree
  sketch: Approximate counting for nlp,'' 2016. [Online]. Available:
  \url{https://arxiv.org/abs/1604.05492}
\BIBentrySTDinterwordspacing

\bibitem{tofino_wedge}
\BIBentryALTinterwordspacing
``Wedge 100bf-32x,'' Edgecore Networks, 2020. [Online]. Available:
  \url{https://bit.ly/2YDeyv2}
\BIBentrySTDinterwordspacing

\bibitem{p4_lan_spec}
\BIBentryALTinterwordspacing
``P4 language and related specifications,'' Open Networking Foundation, 2020.
  [Online]. Available: \url{https://p4.org/specs/}
\BIBentrySTDinterwordspacing

\bibitem{MoshrefYGV16Trumpet}
M.~Moshref, M.~Yu, R.~Govindan, and A.~Vahdat, ``Trumpet: Timely and precise
  triggers in data centers,'' in \emph{ACM SIGCOMM 2016}.\hskip 1em plus 0.5em
  minus 0.4em\relax {ACM}, 2016, pp. 129--143.

\bibitem{li2016flowradar}
Y.~Li, R.~Miao, C.~Kim, and M.~Yu, ``Flowradar: A better netflow for data
  centers,'' in \emph{NSDI 2016}.\hskip 1em plus 0.5em minus 0.4em\relax
  {USENIX} Association, 2016, pp. 311--324.

\bibitem{liu2016one}
Z.~Liu, A.~Manousis, G.~Vorsanger, V.~Sekar, and V.~Braverman, ``One sketch to
  rule them all: Rethinking network flow monitoring with univmon,'' in
  \emph{ACM SIGCOMM 2016}, 2016, pp. 101--114.

\bibitem{metwally2005efficient}
A.~Metwally, D.~Agrawal, and A.~El~Abbadi, ``Efficient computation of frequent
  and top-k elements in data streams,'' in \emph{International conference on
  database theory}.\hskip 1em plus 0.5em minus 0.4em\relax Springer, 2005, pp.
  398--412.

\bibitem{Barefoot_Doc}
\emph{10k Device Family - Switch Architecture Specification}.\hskip 1em plus
  0.5em minus 0.4em\relax Barefoot Networks, 2020.

\bibitem{CAIDA}
\BIBentryALTinterwordspacing
``The cooperative association for internet data analysis, equinix chicago data
  center,'' 2018, [13:00-14:00, Apr 19 2018., from Sao Paulo to New York].
  [Online]. Available: \url{https://www.caida.org}
\BIBentrySTDinterwordspacing

\bibitem{SIMD}
\BIBentryALTinterwordspacing
``Intel streaming simd extensions (sse) technologies,'' Intel, 2020. [Online].
  Available:
  \url{https://software.intel.com/sites/landingpage/IntrinsicsGuide/}
\BIBentrySTDinterwordspacing

\bibitem{hauser2021survey}
F.~Hauser, M.~H{\"a}berle, D.~Merling, S.~Lindner, V.~Gurevich, F.~Zeiger,
  R.~Frank, and M.~Menth, ``A survey on data plane programming with p4:
  Fundamentals, advances, and applied research,'' \emph{arXiv preprint
  arXiv:2101.10632}, 2021.

\bibitem{P4Runtime_API}
\BIBentryALTinterwordspacing
``The p4runtime api,'' P4 Language Consortium, 2021. [Online]. Available:
  \url{https://p4.org/p4-runtime/}
\BIBentrySTDinterwordspacing

\bibitem{jin2018netchain}
X.~Jin, X.~Li, H.~Zhang, N.~Foster, J.~Lee, R.~Soul{\'e}, C.~Kim, and
  I.~Stoica, ``Netchain: Scale-free sub-rtt coordination,'' in \emph{NSDI
  2018}, 2018, pp. 35--49.

\bibitem{jin2017netcache}
X.~Jin, X.~Li, H.~Zhang, R.~Soul{\'e}, J.~Lee, N.~Foster, C.~Kim, and
  I.~Stoica, ``Netcache: Balancing key-value stores with fast in-network
  caching,'' in \emph{SOSP 2017}, 2017, pp. 121--136.

\bibitem{sivaraman2017heavy}
V.~Sivaraman, S.~Narayana, O.~Rottenstreich, S.~Muthukrishnan, and J.~Rexford,
  ``Heavy-hitter detection entirely in the data plane,'' in \emph{SOSR 2017},
  2017, pp. 164--176.

\bibitem{peterson1972error}
W.~W. Peterson, W.~Peterson, E.~Weldon, and E.~Weldon, \emph{Error-correcting
  codes}.\hskip 1em plus 0.5em minus 0.4em\relax MIT press, 1972.

\bibitem{miao2017silkroad}
R.~Miao, H.~Zeng, C.~Kim, J.~Lee, and M.~Yu, ``Silkroad: Making stateful
  layer-4 load balancing fast and cheap using switching asics,'' in \emph{ACM
  SIGCOMM 2017}, 2017, pp. 15--28.

\bibitem{libtins_pkt_gen}
M.~Fontanini, ``libtins: Packet crafting and sniffing library,''
  http://libtins.github.io/, 2020.

\bibitem{MoshrefYGV14DREAM}
M.~Moshref, M.~Yu, R.~Govindan, and A.~Vahdat, ``{DREAM:} dynamic resource
  allocation for software-defined measurement,'' in \emph{ACm SIGCOMM
  2014}.\hskip 1em plus 0.5em minus 0.4em\relax {ACM}, 2014, pp. 419--430.

\bibitem{GuoYXDHMLWPCLK15Pingmesh}
C.~Guo, L.~Yuan, D.~Xiang, Y.~Dang, R.~Huang, D.~A. Maltz, Z.~Liu, V.~Wang,
  B.~Pang, H.~Chen, Z.~Lin, and V.~Kurien, ``Pingmesh: A large-scale system for
  data center network latency measurement and analysis,'' in \emph{ACM SIGCOMM
  2015}.\hskip 1em plus 0.5em minus 0.4em\relax {ACM}, 2015, pp. 139--152.

\bibitem{NarayanaSNGAAJK17Marple}
S.~Narayana, A.~Sivaraman, V.~Nathan, P.~Goyal, V.~Arun, M.~Alizadeh,
  V.~Jeyakumar, and C.~Kim, ``Language-directed hardware design for network
  performance monitoring,'' in \emph{ACM SIGCOMM 2017}.\hskip 1em plus 0.5em
  minus 0.4em\relax {ACM}, 2017, pp. 85--98.

\end{thebibliography}

\ifCLASSOPTIONcaptionsoff
  \newpage
\fi



%


\newpage

%








\end{document}